\documentclass[twocolumn,trackchanges]{aastex63}

\pdfoutput=1
\usepackage[whole]{bxcjkjatype}
\usepackage{amsmath,gensymb}
\usepackage{graphicx}
\graphicspath{ {./fig/} }
\usepackage{bm}
\usepackage{makecell}
\newcommand{\tabincell}[2]{\begin{tabular}{@{}#1@{}}#2\end{tabular}}
\newcommand{\intd}{\text{d}}

\newcommand{\h}{\tfrac{1}{2}}
\newcommand{\second}{\,\mathrm{s}}
\newcommand{\au}{\,\mathrm{AU}}
\newcommand{\kyr}{\,\mathrm{kyr}}
\newcommand{\solarmass}{\,\mathrm{M}_\sun}

\begin{document}
\title{Magnetic Spirals in Accretion Flows Originated from Misaligned Magnetic Field}

\author{Weixiao Wang}
\affiliation{Shanghai Astronomical Observatory, Chinese Academy of Sciences, 80 Nandan Road, Shanghai 200030, People's Republic of China}
\affiliation{University of Chinese Academy of Sciences, 19A Yuquan Road, Beijing 100049, People's Republic of China}
\affiliation{Academia Sinica, Institute of Astronomy and Astrophysics, Taipei, Taiwan}

\author[0000-0002-8782-4664]{Miikka S. V\"ais\"al\"a}
\affiliation{Academia Sinica, Institute of Astronomy and Astrophysics, Taipei, Taiwan}

\author[0000-0001-8385-9838]{Hsien Shang}
\affiliation{Academia Sinica, Institute of Astronomy and Astrophysics, Taipei, Taiwan}

\author[0000-0001-5557-5387]{Ruben Krasnopolsky}
\affiliation{Academia Sinica, Institute of Astronomy and Astrophysics, Taipei, Taiwan}

\author{Zhi-Yun Li}
\affiliation{Astronomy Department, University of Virginia, Charlottesville, VA 22904, USA}

\author[0000-0003-3581-1834]{Ka Ho Lam}
\affiliation{Academia Sinica, Institute of Astronomy and Astrophysics, Taipei, Taiwan}
\affiliation{Astronomy Department, University of Virginia, Charlottesville, VA 22904, USA}

\author{Feng Yuan}
\affiliation{Shanghai Astronomical Observatory, Chinese Academy of Sciences, 80 Nandan Road, Shanghai 200030, People's Republic of China}

\correspondingauthor{Weixiao Wang, Miikka S. V\"ais\"al\"a, Hsien Shang}
\email{wxwang@shao.ac.cn, mvaisala, shang@asiaa.sinica.edu.tw}

\begin{abstract}
Misalignment between rotation and magnetic field has been suggested to be one type of physical mechanisms which can easen the effects of magnetic braking during collapse of cloud cores leading to formation of protostellar disks. However, its essential factors are poorly understood. Therefore, we perform a more detailed analysis of the physics involved. We analyze existing simulation data to measure the system torques, mass accretion rates and Toomre Q parameters. We also examine the presence of shocks in the system. While advective torques are generally the strongest, we find that magnetic and gravitational torques can play substantial roles in how angular momentum is transferred during the disk formation process. Magnetic torques can shape the accretion flows, creating two-armed magnetized inflow spirals aligned with the magnetic field. We find evidence of an accretion shock that is aligned according to the spiral structure of the system. Inclusion of ambipolar diffusion as explored in this work has shown a slight influence in the small scale structures but not in the main morphology. We discuss potential candidate systems where some of these phenomena could be present.
\end{abstract}


\section{Introduction}\label{sec:intro}

Since Atacama Large Millimeter Array (ALMA) has begun its observations, we have entered into a new era of understanding circumstellar disks. Following the publication of first ALMA continuum observations of HL Tau \citep{ALMA2015}, there has been fast accumulation of observations of circumstellar disks at various stages of their evolution, including protostellar stage. DSHARP \citep{DHARPI2018} survey has completed, and there have been several individual observations from nearby regions such as Chamaeleon I \citep{Long2017, Long2018a}, Corona Australis \citep{Cazzoletti2019}, IC 348 \citep{ruiz2018}, Lupus \citep{Ansdell2016, Ansdell2018, Tazzari2020}, Lynds 1641 \citep{Grant2021}, Ophiuchus \citep{Cieza2021, Sadavoy2019}, Orion \citep{Ansdell2017, Dutta2020, Hsu2020, Tobin2019, Tobin2020, Sahu2021, Sheehan2020}, Perseus \citep{Yang2021}, Upper Scorpius OB1 \citep{Carpenter2014}, Taurus \citep{Long2018, Long2019, Podio2020}.

However, we have still very limited knowledge about the early protostellar stages of circumstellar disks. Observations of these very early stages of disks formation is challenging because such objects are surrounded by envelope of the collapsing core. The envelope can make it very difficult to differentiate various chemical and physical properties as distinct localized features \citep[see e.g.][and the references therein]{Harsono2021}.  

Theoretical problems are likely not easier either. Disk formation can be affected by the magnetic field, dynamical properties of a prestellar core, chemical and other microphysical effects. 
Ideal magnetohydrodynamics (hereafter MHD) assumptions also lead into magnetic braking catastrophe preventing disk formation \citep{ALS2003}. 

As explained in \citet{GLSA2006}, newly formed stars from perfectly ideal MHD scenario would have an excessively strong initial magnetic field compared to what is observed around T Tauri stars \citep[see e.g.][]{mestel1956, JohnsKrull2004}. 
We have to find a way to deal with the magnetic braking problem, but it is impossible to model everything at once, and therefore we choose to focus on the inclined magnetic field and, in a supplementary manner, the influence of one of the non-ideal MHD effects.

One possible way of making a rotationally supported disk (hereafter RSD) in an ideal MHD system is by misalignment between the mean magnetic field and rotation axis. This was demonstrated by \citet{Hennebelle2009} and \citet{JHC2012}, and further elaborated by \citet{LKS2013}. \citet{LKS2013} found in agreement with previous work that with a sufficiently large mass-to-flux ratio, disk formation became a possibility. \citet{MV2019} further analyzed the data of \citet{LKS2013} with a simple radiative transfer approach, and in addition to highlighting physical effects that were missed in the original study, they noted the importance of spiral patterns, especially with the spiralling inflows, as a potential observational signature of misalignment based on disk formation. \citet{MV2019} also noted that the disk can be prone to non-circular motions leading to spiral perturbations and that with sufficient magnetic field, rings could form. However, they did not do full analysis of torques.

Spirals during star formation can emerge due to gravitational disturbances and gravitational instability \citep[see e.g.][and the references therein]{Kratter2016}. Therefore, it is imperative to examine how magnetic and gravitational forces are acting during misaligned collapse, to determine both how misaligned collapse can work and which factors are decisive. This paper focuses on the magnetized spirals that originate from the misaligned magnetic field; previous works in this line of study had a different focus (\citealp{LKS2013} focused on the presence of RSD structures, \citealp{MV2019} focused on finding potential observable features).

Understanding the mechanisms of magnetic spirals can have significant implications. 
With recently observed streamers into the Class 0 object Per-emb-2 \citep{Pineda2020}, the arc seen by \citet{Grant2021} in [MGM2012] 512, and the field configuration in HH211 \citep{Lee2019}, there are observational possibilities to examine spiralling inflows.
\citet{Sanhueza2021} have also been able to demonstrate streamers in massive star formation environment of IRAS 18089--1732.
Spiral behaviour can also happen in a disk plane, like in HH 111 VLA 1 as observed by \citet{Lee2020}. 
As further elaborated in this study, magnetic field can cause formation of warped protostellar rings, which can provide one possible explanation of some observed warped disks like L1527 IRS around IRAS 04368+2557 \citep{Sakai2019,Nakatani2020}.

The paper is organized as follows. In Section \ref{sec:methods}, we present the central analysis methods that we utilize. In Section \ref{sec:results} we display the results of our analysis. In Section \ref{sec:disc} we discuss implication of our results especially with respect to observations, and in Section \ref{sec:summary} we summarize the paper.

\section{Methods}\label{sec:methods}

In this paper we perform further analysis and post-processing on data first published in \citet{LKS2013}.
Using the non-ideal MHD code Zeus-TW \citep{KLS2010} and incorporating self-gravity, the authors performed MHD simulations of an initially rotating, uniform, dense core with misaligned magnetic field in a spherical coordinate system. The initial magnetic field was tilted uniformly away from the rotation axis by different angles. The grid is non-uniform in the radial and meridional directions, with constant ratios ($\sim 1.08$ for both $r$ and $\theta$) between the non-uniform widths of adjacent active zones. The grid is uniform in the azimuthal direction. Implemented parameters, initial conditions and boundary conditions are shown in Table \ref{tab:par}. The tilt angle $\theta_0$ and the dimensionless mass-to-flux ratio $\lambda$ are major free parameters in their series of models. To prevent numerical ``hot zones'', meaning the locations in the numerical domain where the Alfv\'en velocity becomes extremely large leading to infinitesimal time scale, from halting the simulations, a small, spatially uniform resistivity $\eta=10^{17}{\rm cm^2\,s^{-1}}$ was implemented. The ``hot zones'' arose at the start of the rapid accretion, and therefore resistivity was applied after the central mass grew up to about $10^{-7}\solarmass$. Models of reduced resistivity ($\eta=10^{16}\,{\rm cm^2\,s^{-1}}$ and $\eta=0$) verified that $\eta=10^{17}\,{\rm cm^2\,s^{-1}}$ is small enough to be insignificant for the sake of numerical improvement.

\begin{deluxetable*}{llll}[htb!]
\tablenum{1}
\tablecaption{Simulation setup\label{tab:par}}
\tablewidth{\textwidth}
\tablehead{
\colhead{Initial} & \colhead{Boundary} & \colhead{Grid Setup} &
\colhead{Equation of state}
}
\startdata
\tabincell{p{0.35\textwidth}}{Uniform density $\rho_0=4.77\times10^{-19}{\rm g\,cm^{-3}}$;\\
Uniform field $B_0(\lambda)$;\\
Solid-body rotation speed $\Omega_0=10^{-13}\,{\rm s^{-1}}$} &
\tabincell{p{0.1\textwidth}}{${r}:$ outflow\\
${\theta}:$ reflective\\
${\phi}:$ periodic} & \tabincell{p{0.2\textwidth}}{Mesh: $96\times64\times60$\\
${r}:$ $10^{14}-10^{17}$cm;\\
${\theta}:$ $0-\pi$;\\
${\phi}:$  $0-2\pi$} & \tabincell{p{0.22\textwidth}}{Isothermal (with a sound speed $a=0.2{\rm\, km\,s^{-1}}$) for $\rho<\rho_c(\equiv10^{-13}\,\rm{g}\,\rm{cm}^{-3})$, polytropic ($p\propto\rho^{5/3}$) for $\rho>\rho_c$.} \\
\enddata
\tablecomments{From \citet{LKS2013}. The initial spherical core with uniform density has $1\solarmass$ in mass and about $6685\au$ ($10^{17}\rm{cm}$) in radius, corresponding to a free-fall time of about $95.1\kyr$ ($3\times10^{12}\second$).}
\end{deluxetable*}

In an effort to characterize the observable properties of magnetically misaligned protostellar disk systems, \citet{MV2019} identified distinguishable visual features of each model. 
Especially noteworthy were the spiral patterns. 
To further quantitatively study them, in this paper we choose models with various types of spirals and models without spiral patterns for comparison. The models of interest are listed in Table \ref{tab:models}. Model G, which has robust RSD and clear spiral structure, is specified as the fiducial model.

\begin{deluxetable*}{cccccp{0.3\columnwidth}}
\tablenum{2}
\tablecaption{Selected models and results\label{tab:models}}
\tablewidth{\columnwidth}
\tablehead{
\colhead{Model} & \colhead{$\lambda$} & \colhead{$\theta_0$} &
\colhead{RSD} & \colhead{Visual type} & \colhead{Dominant torque in the central region at late stage}}
\startdata
G & 9.72 & 90 & Yes/robust & Clear Spiral & Gravitational torque \\
H & 4.86 & 90 & Yes/porous & Leaking Spiral & Magnetic torque\\
I & 2.92 & 90 & No & Looped Axis & Magnetic torque \\
A & 9.72 & 0  & No & Looped Plane & Magnetic torque  \\
D & 9.72 & 45 & Yes/porous & Looped Plane / Leaking Spiral & Magnetic torque\\
Galpha & 9.72 & 90 & Yes/robust & Clear Spiral & Gravitational torque\\
\enddata
\tablecomments{The second to the fifth columns of parameters of the first four rows of models are from \citet{LKS2013}, with the classified visual types from \citet{MV2019}. The last row of model Galpha is a new case with ambipolar diffusion (see details in Section \ref{sec:AD}). The last column lists the results of dominant torques.}
\end{deluxetable*}

\subsection{Calculating torques}\label{sec:mthd_torq}

Analyzing angular momentum transfer of the disk is important for investigating the dynamics of disk formation and its spiral pattern. The key questions are how much the angular momentum is conserved, how the angular momentum is redistributed and which mechanism dominates in this process.

In an Eulerian frame conservation of angular momentum in the case of magnetized inviscid flow is given by
\begin{align}
	\frac{\partial \mathbf{J}}{\partial t} = & -\int_S(\rho\mathbf{r}\times\mathbf{v})\mathbf{v}\cdot\text{d}\mathbf{S}\\
	& + \frac{1}{4\pi}\int_V\mathbf{r}\times((\nabla\times\mathbf{B})\times\mathbf{B}))\,\text{d}V \nonumber\\
	& - \int_V \mathbf{r}\times\nabla p\,\intd V + \int_V \rho\mathbf{r}\times\mathbf{g}\,\intd V \nonumber\ ,
\end{align}
where $\mathbf{J}=\int_V\rho\mathbf{r}\times\mathbf{v}\,\text{d}V$ is the angular momentum within a control volume $V$, and $\mathbf{g}=-\nabla\Phi$ is the gravitational force per unit mass per unit volume. The terms on the right hand side (hereafter RHS) are torques due to advection, magnetic field, thermal pressure, and gravity. 

To better demonstrate the mechanisms of disk rotation, we calculate in spherical coordinates the z-component of the conservation of angular momentum
\begin{align}\label{eqn:jz}
  \frac{\partial J_z}{\partial t} = & -\int_S(\rho r\sin{\theta} v_\phi)(\mathbf{v}\cdot\text{d}\mathbf{S})\\
  & + \frac{1}{4\pi}\int_S (\rho r\sin{\theta} B_\phi)(\mathbf{B}\cdot\text{d}\mathbf{S}) \nonumber\\
  & - \frac{1}{4\pi}\int_V \frac{\partial}{\partial\phi}(\frac{B^2}{2})\intd V - \int_V \frac{\partial p}{\partial\phi}\intd V  - \int_V \rho \frac{\partial\Phi}{\partial\phi} \intd V \nonumber\ ,
\end{align}
which contains the torques relative to the z-axis. The integration control volume $V$ and its surface $S$ are usually chosen as the volume $V_\text{cell}$ and surface $S_\text{cell}$ of a grid cell. When torques are calculated locally, the pressure terms cannot be neglected, which is not the case for the integral within a sphere or a cylinder. In order to compare local values from cell to cell, we divide torques by the cell volume.

We also calculate the specific angular momentum defined as $\mathbf{J}_{\rm{spc}}=\mathbf{J}/\int_{V_{\rm{cell}}} \rho\intd V$ as well as its z-component $J_{\rm{spc,z}}$, and plot its spatial distribution.

\subsection{Detecting gravitational instability}\label{sec:mthd_gi}

If a gaseous disk is massive enough, it can be gravitationally unstable.
Following \citet{MV2019} remarks on the potential influence of self-gravity, here we investigate the role of gravitational effects on substructures.

The parameter describing gravitational instability is $Q\equiv\frac{a\kappa}{\pi G\Sigma}$ \citep{LinShu1964, Toomre1964,Goldreich1965}, where $a$ is the sound speed, $\kappa$ is the epicyclic frequency defined as $\kappa^2\equiv 4\Omega^2+2R\Omega\frac{\partial\Omega}{\partial R}$ ($\kappa=\Omega$ for Keplerian rotation), and $\Sigma=\int^z_{-z}\rho \intd z$ is the disk surface density.
In this calculation, $\Sigma$ will be obtained from PERSPECTIVE \footnote{PERSPECTIVE is a light-weight radiative transfer code for the purpose of examining observational variables, or other integrable quantities, based on simulations. For a more complete description of PERSPECTIVE, see Section 3 in \citet{MV2019}.} by integrating system along the line of sight via interpolated ray-tracing method, as in \citet{MV2019}. 

The modification of Toomre $Q$ parameter for magnetized disk can be written as $Q_M\equiv \frac{a^\prime\kappa}{\pi G\Sigma\epsilon}$ in general \citep{SL1997,LGCA2010}, where $a^\prime$ is the isothermal magnetosonic speed and $\epsilon=1-\lambda^{-2}$ is a function of mass-to-flux ratio $\lambda$. With the consideration of magnetic field, the critical limit $Q_M$ becomes greater than the original Toomre $Q$ \citep{LGCA2010}. When non-ideal MHD effects are included, the limit can be reduced to that of the hydrodynamic case \citep{DB2021}. 
However for our simple purposes, we neglect non-ideal MHD effects and only take the hydrodynamical case (lower limit). As shown in Section \ref{sec:gi}, it is sufficiently informative for our purposes.

\subsection{Shock Identification}\label{sec:mthd_shock}

Shock waves, such as accretion shocks onto a disk, are salient features and can play a significant role in influencing the dynamics of the accretion flow and thereby the disk \citep[e.g.][]{YBL1993,YBL1995,YB1999}.

Accurately and completely detecting shocks in numerical simulations can be difficult.
In spite of that, when shock properties are not our focus, a simple method to find their location would be adequate. The algorithm for detecting shocks implemented in this paper is based on two threshold conditions: (1) thermal pressure gradient larger than an empirical threshold; and (2) positive convergence of velocity ($-\nabla\cdot\mathbf{v}$) larger than another empirical threshold.
These thresholds filter out discontinuities that are too weak to detect and artificial jumps due to numerical deviations.

\subsection{Characterizing spiral structure}\label{sec:mthd_spiral}

As shown in Table \ref{tab:models}, spiral structures are present together with RSD in visible features. \citet{LKS2013} only explored the two-armed ``pseudo-spirals'' (analogously referring to ``pseudodisk'') which are part of a magnetically-induced curtain. By looking into time evolution of the disk formation, \citet{MV2019} noticed inner spirals through morphology. However, in these two articles more detailed properties of the spirals have not been quantitatively further studied.

To characterize spiral structure, we manually fit the 2D column density map by a simple logarithmic form given by
\begin{equation}
    \varphi = b\,\rm{ln}r+const.
\end{equation}
where $(r,\varphi)$ denotes the polar coordinate and $(b,const)$ is the parameter space to explore. The corresponding pitch angle would be
\begin{equation}
    \alpha = {\rm arctan}\frac{1}{b},
\end{equation}
which does not depend on $const$.

\section{Results}\label{sec:results}
In this section, results of selected models are presented in detail. Spatial distribution of angular momentum and torques, including the dominant torque, are described in Section \ref{sec:rst_torq};
the role of gravitational instability is studied in Section \ref{sec:gi};
the interplay of gravitational and magnetic effects in the spirals is shown in \ref{sec:spirals};
trends in time for central mass and angular momentum are explored in \ref{sec:growth};
results of shock-detection are shown in \ref{sec:shocks}; the warped disk and rings present in the precessing model H are studied in \ref{sec:warped}; and the dependence on ambipolar diffusion is demonstrated in \ref{sec:AD}.

Some of the phenomena were already discovered in the \citet{LKS2013} and \citet{MV2019}. However, due to their different focus, their analysis was either limited or merely their existence was noted without more quantitative analysis.

\subsection{Angular Momentum and Torques}\label{sec:rst_torq}
\begin{figure*}[htb]
	\centering
	\includegraphics[width=\textwidth]{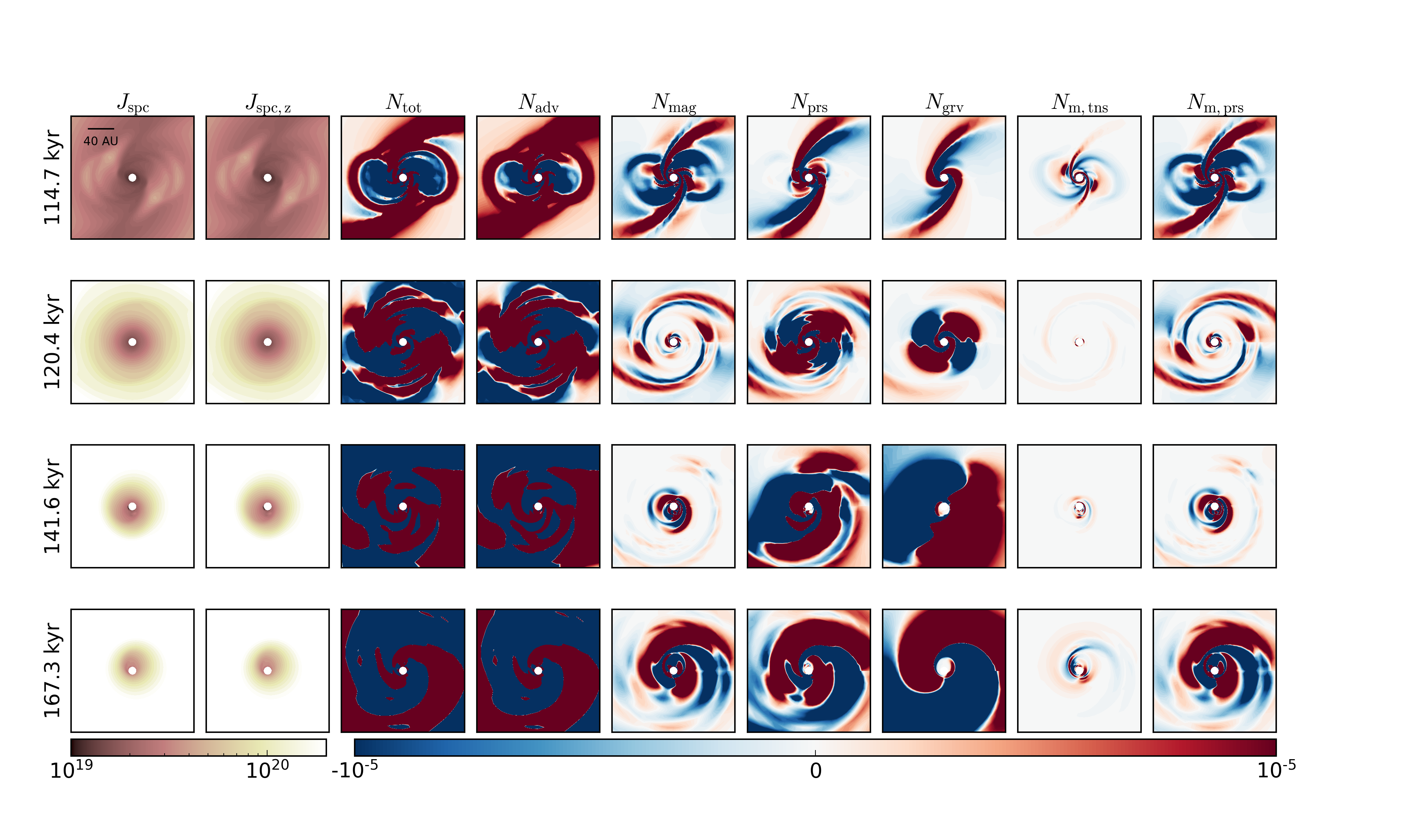}
	\includegraphics[width=\textwidth]{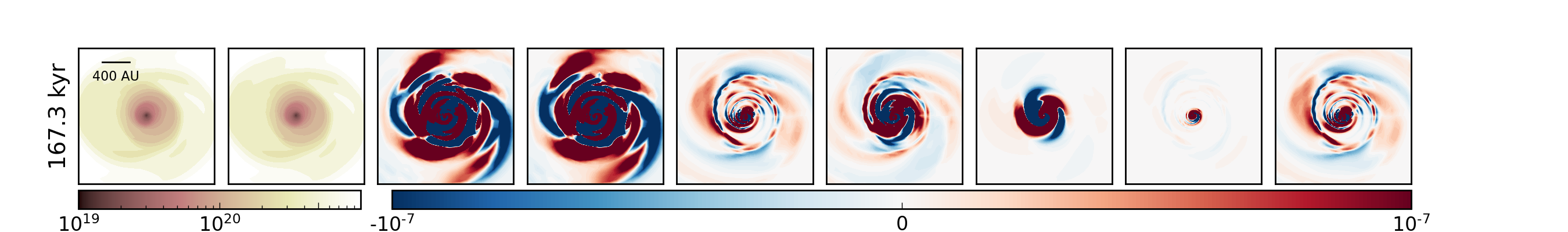}
	\caption{Color map of specific angular momentum and torques of model G in time sequence. Panels from left to right are the specific angular momentum per unit volume $J_{\rm{spc}}$, the z-component of the specific angular momentum per unit volume $J_{\rm spc,z}$, the total torque $N_{\rm{tot}}$, the advective torque $N_{\rm{adv}}$, the magnetic torque $N_{\rm{mag}}$, the torque due to pressure gradient $N_{\rm{prs}}$, the gravitational torque $N_{\rm{grv}}$, the torque due to magnetic tension $N_{\rm{m,tns}}$, and the torque due to magnetic pressure $N_{\rm{m,prs}}$, respectively. First four rows are results in the same color scale at different times in sequence. The box size is $200\times200\,\mathrm{\au^2}$. The last row displays results at the same frame as the fourth row but in different color scale, and the box size is $2000\times2000\,\mathrm{\au^2}$. }\label{fig:g5}
\end{figure*}

\begin{figure*}[htb]
	\centering
	\includegraphics[width=\textwidth]{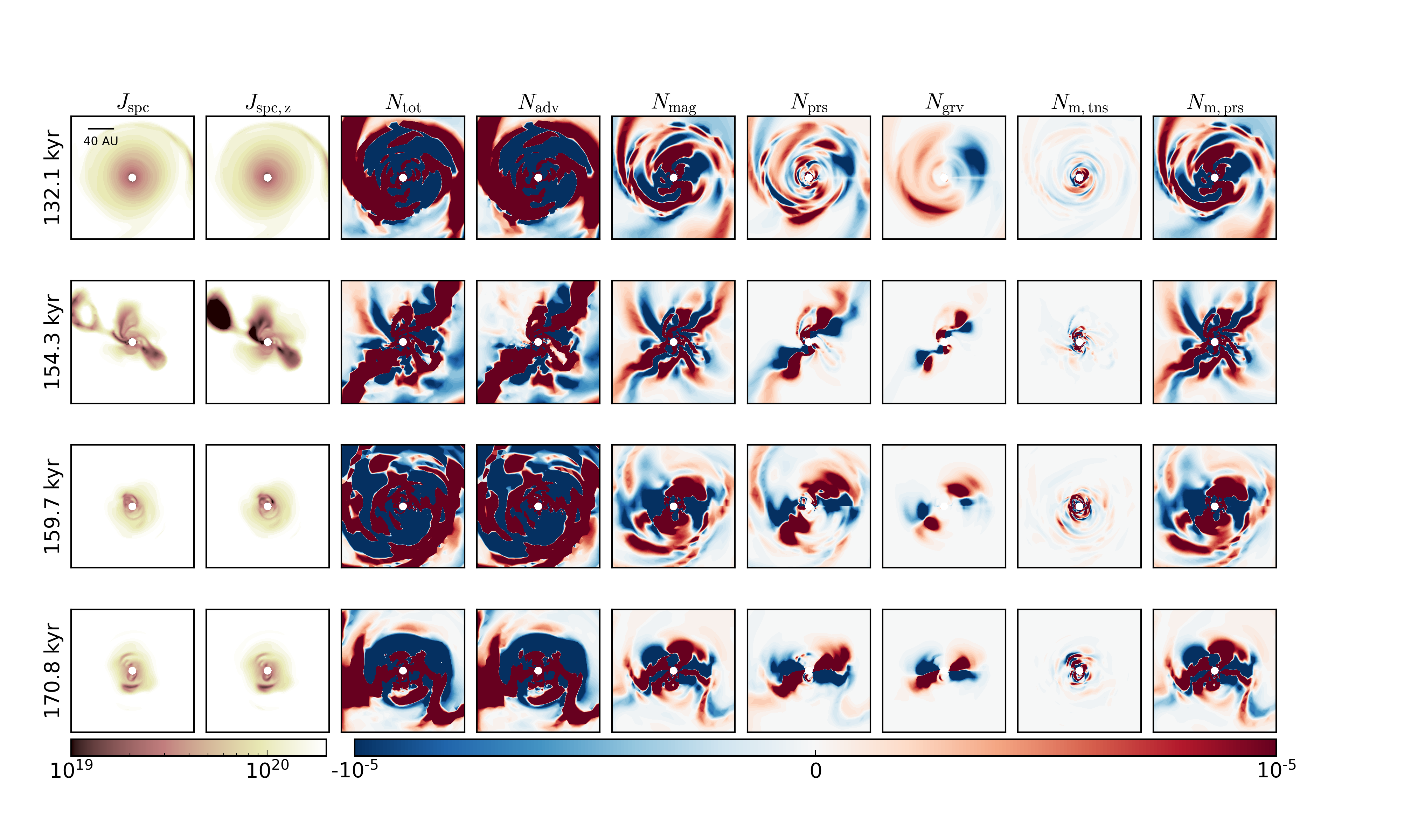}
	\includegraphics[width=\textwidth]{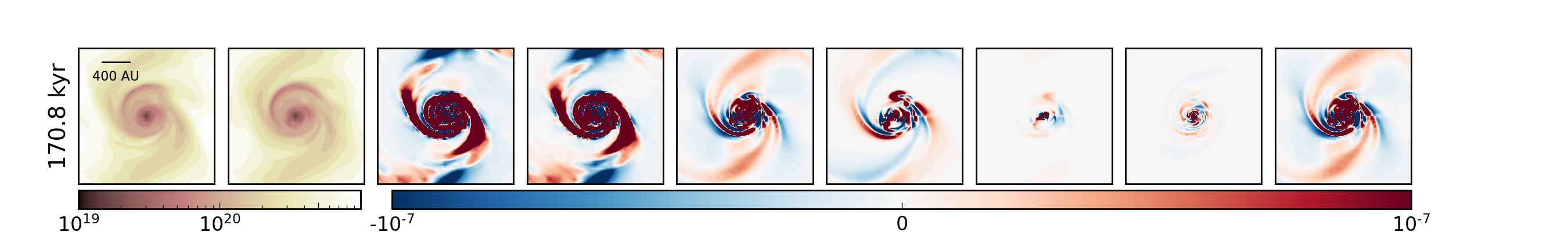}
	\caption{Same as Figure \ref{fig:g5} but for model H.}\label{fig:h5}
\end{figure*}

\begin{figure*}[htb]
	\centering
	\includegraphics[width=\textwidth]{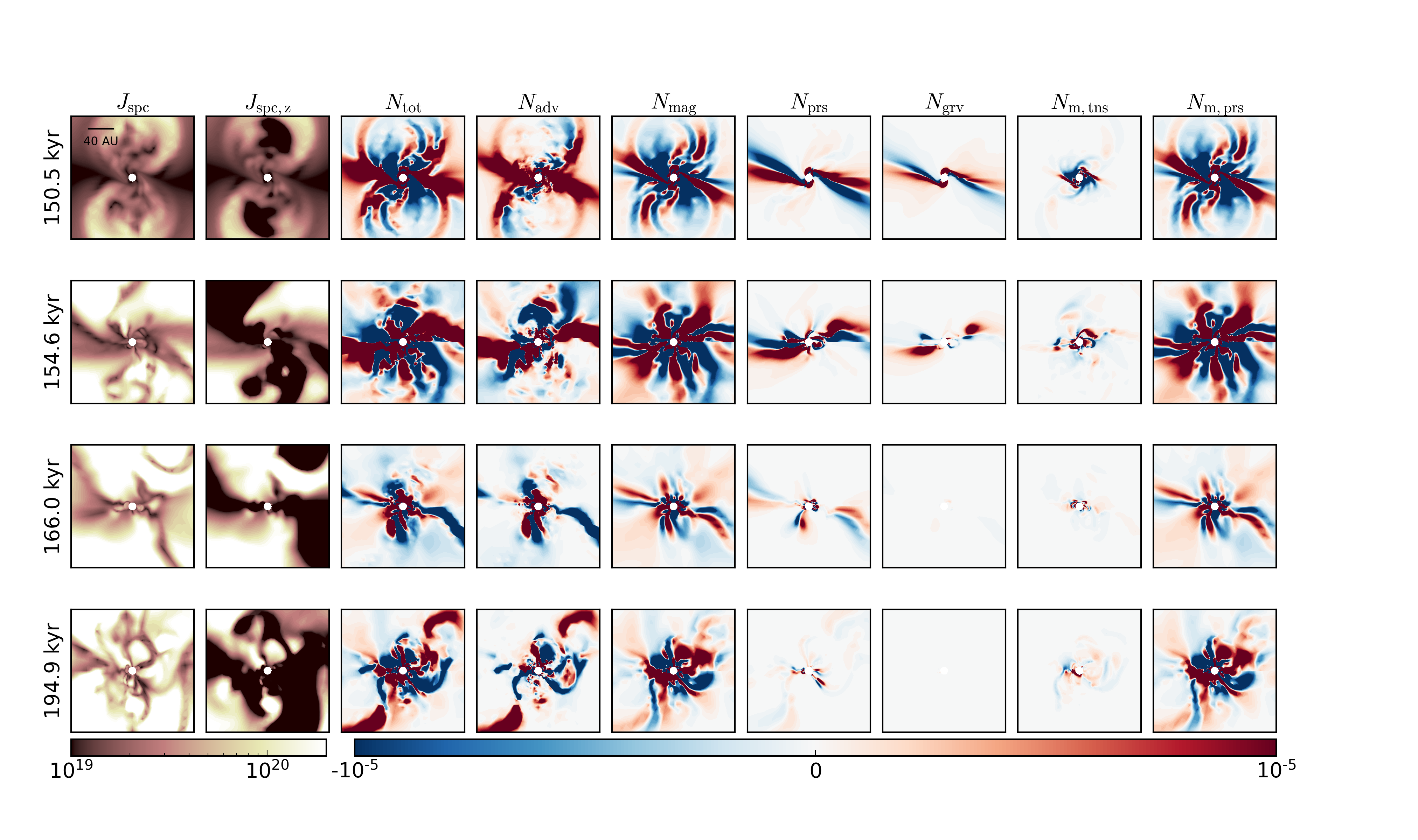}
	\includegraphics[width=\textwidth]{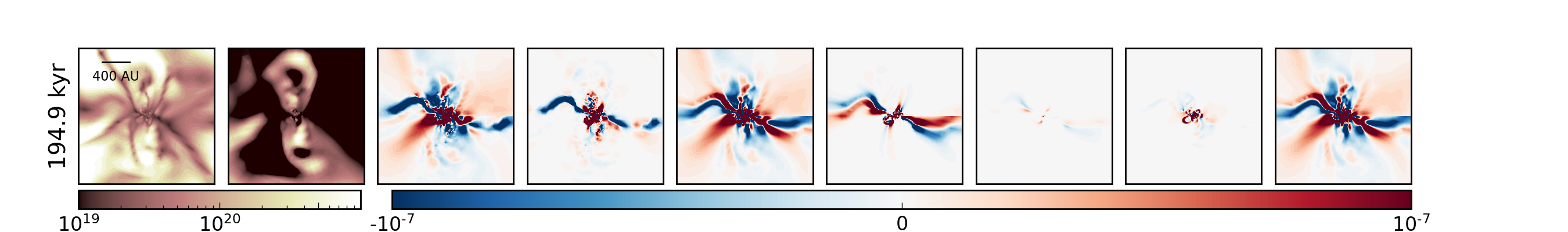}
	\caption{Same as Figure \ref{fig:g5} but for model I}\label{fig:i5}
\end{figure*}

\begin{figure*}[htb]
	\centering
	\includegraphics[width=\textwidth]{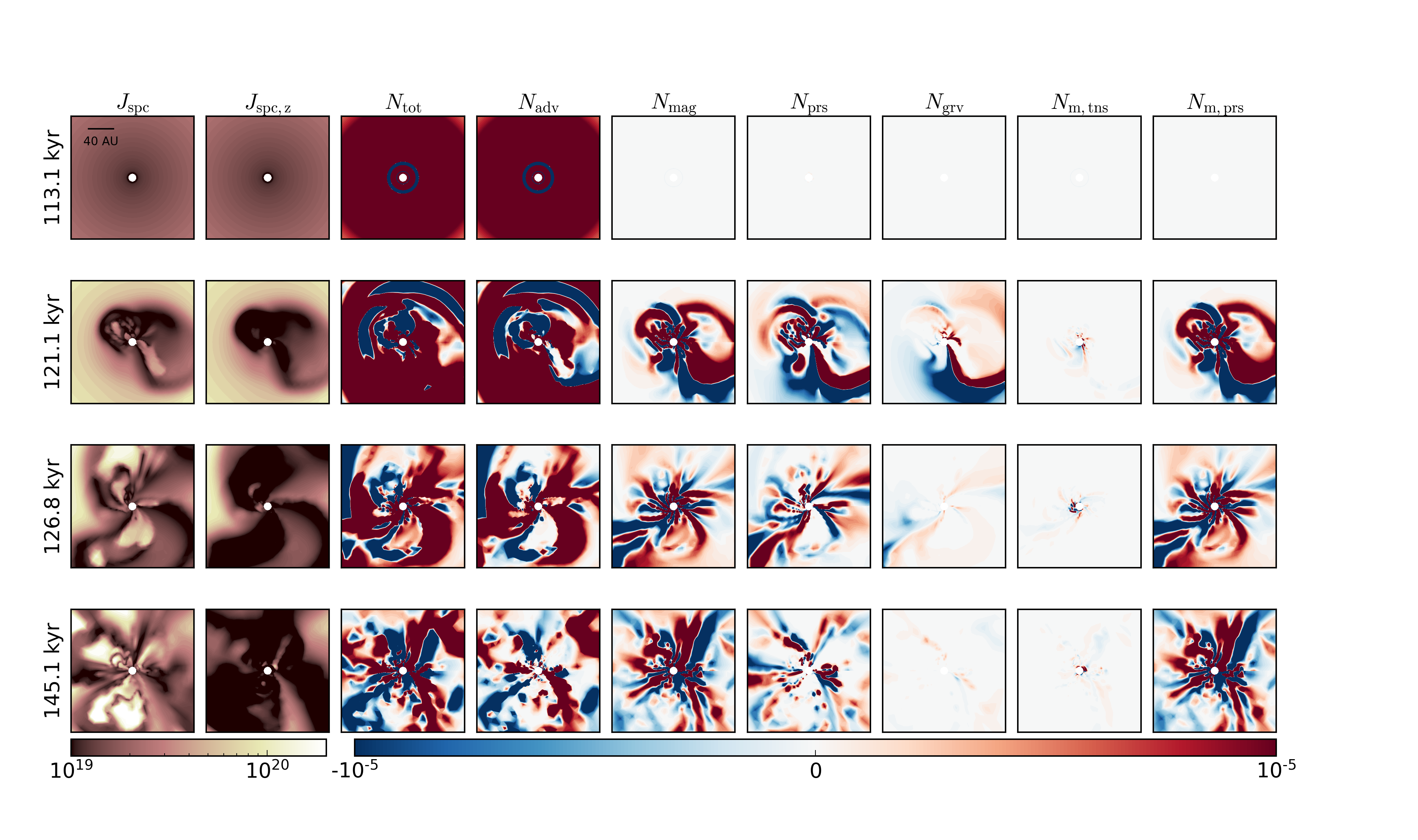}
	\includegraphics[width=\textwidth]{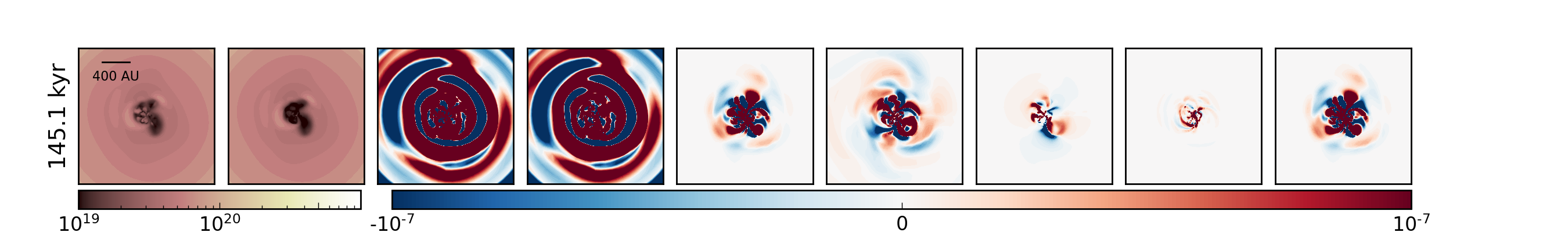}
	\caption{Same as Figure \ref{fig:g5} but for model A}\label{fig:a5}
\end{figure*}

\begin{figure*}[htb]
	\centering
	\includegraphics[width=\textwidth]{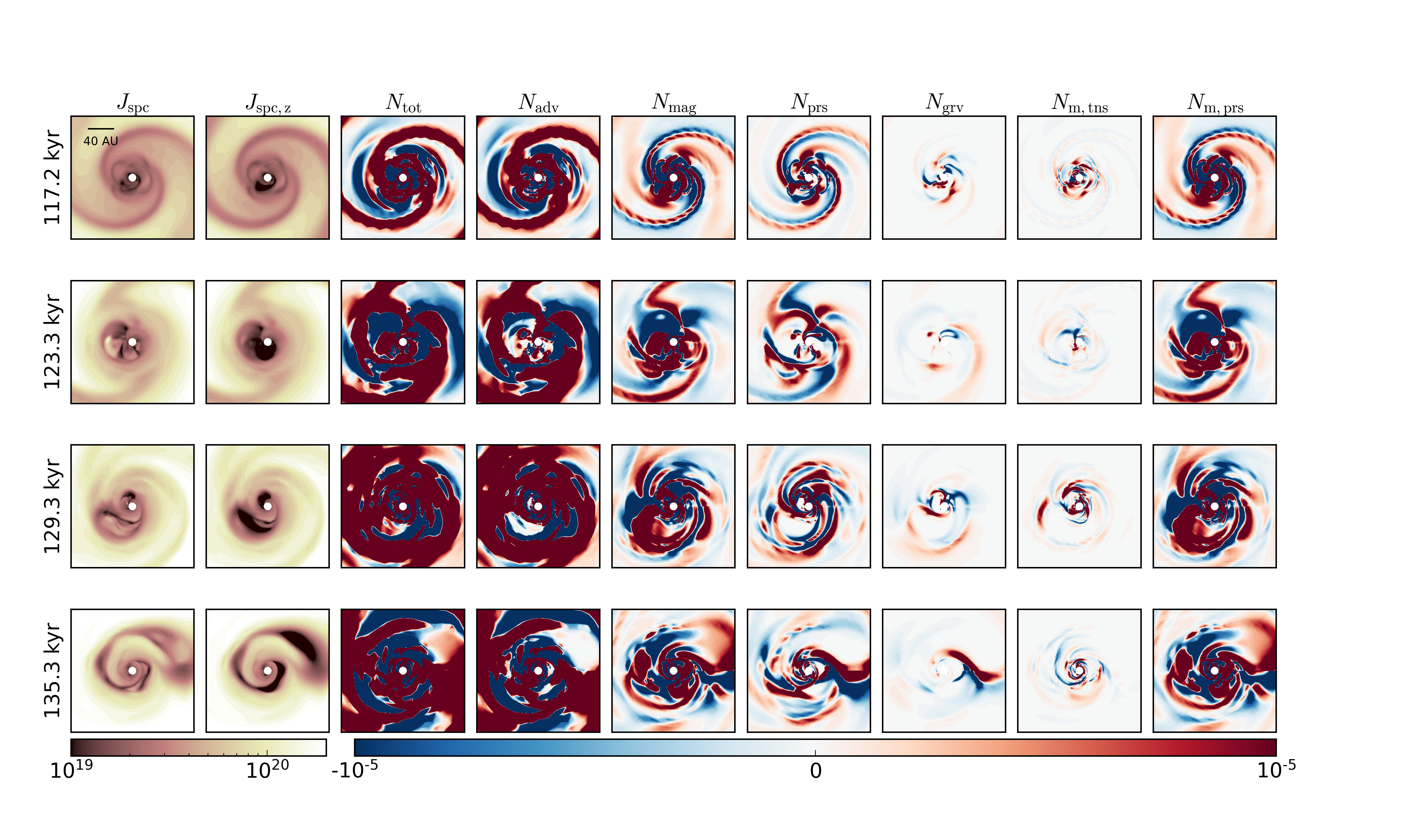}
	\includegraphics[width=\textwidth]{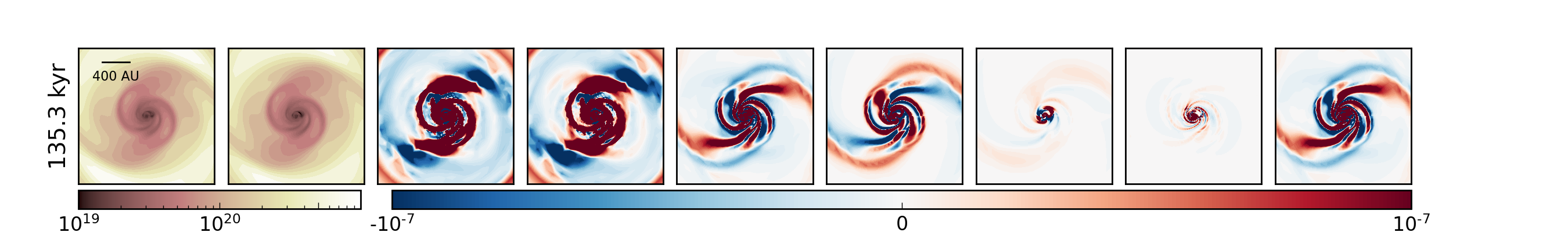}
	\caption{Same as Figure \ref{fig:g5} but for model D}\label{fig:d5}
\end{figure*}

\begin{figure*}[htb]
	\centering
	\includegraphics[width=0.7\textwidth]{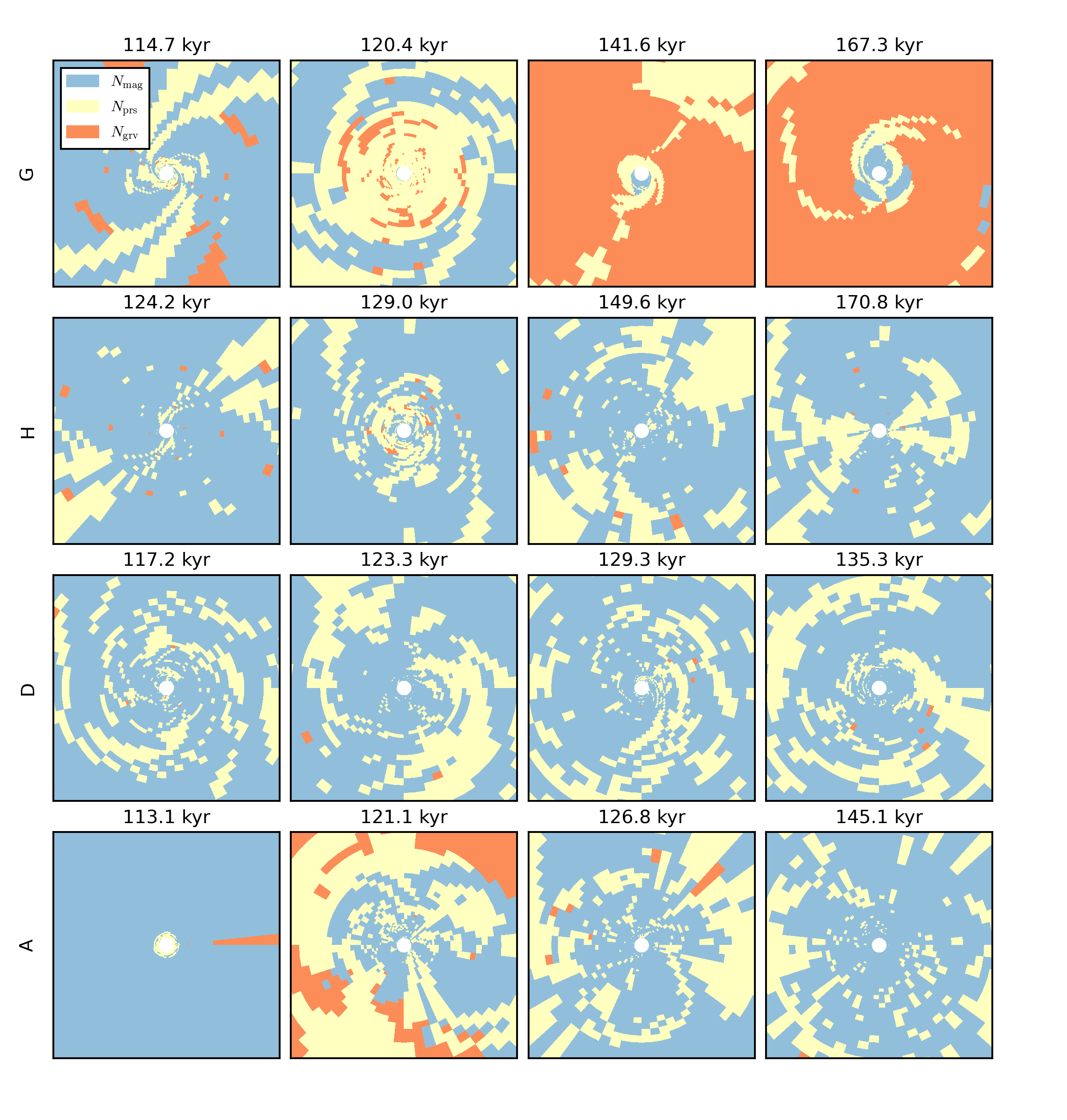}
	\caption{Panels from top to bottom indicate in color the dominant terms of models G, H, D and A: magnetic torque in blue gray, pressure gradient torque in light yellow and gravitational torque in orange. Each panel range extends from $-100\au$ to $100\au$ in both x and y axis. Model I shows similar results here to model H, so it is not shown here.}\label{fig:domTerm}
\end{figure*}

Torques and angular momentum transfer play a central role in protostellar disk formation. Figures from \ref{fig:g5} to \ref{fig:d5} show the angular momentum and torques at midplane\footnote{The first layer of gridpoints above the midplane.} within $100\au$ for their respective models in time sequence. Figures \ref{fig:g5}, \ref{fig:h5}, \ref{fig:i5}, \ref{fig:a5}, and \ref{fig:d5} respectively feature models G, H, I, A, and D.

Models G, H, and I have the same misalignment (largest angle) but different mass-to-flux ratios. As the magnetic field is stronger (smaller mass-to-flux ratio), the specific angular momentum is less preserved and its spatial distribution becomes more chaotic and less azimuthally symmetrical.

Spirals in density and magnetic fields are characteristic features for models with misalignment between the mean magnetic field and the rotation axis.
In such models (G, H, I), 
shown in Figures \ref{fig:g5}-\ref{fig:i5}, a spiral structure as well as a ring-like structure appears in torques (see the first row of Figures \ref{fig:g5}-\ref{fig:i5}). 
The large spirals at early stage  
correspond to the spirals in density along which the flow is falling in. The total torque along spirals is positive, which indicates the flow gains angular momentum through the spirals. 

As major accretion begins, the large spirals are interfered by either the rapid rotation with the presence of disk (model G and H) 
or braking of almost all rotation (model I). In contrast, no spiral structure appears in the aligned model (A, see Figure \ref{fig:a5}), but instead the disk plane becomes dominated by decoupling-enabled magnetic structures \citep[DEMS.][]{ZLNKS2011}, which manifest as low-density region evacuated by decoupled magnetic flux near the protostar.

As for partially misaligned model (D, see Figure \ref{fig:d5}), the large-scale spiral structure in torques and angular momentum shows up at early stage as well. The low density ``holes'', which are magnetically dominated DEMS, have low angular momentum. Though it has the same initial magnetic field strength as model G, partial misalignment cannot completely suppress the formation of DEMS. Also, symmetric pattern in thermal pressure gradient torque and gravitational torque in total misaligned models (G, H, I) disappears in model D. 

Locally, i.e.\ calculated on a single grid cell, the magnetic pressure gradient torque can be more significant than magnetic tension torque so that the former cannot be neglected. Globally, i.e.\ calculated by integrating over a sphere, the magnetic pressure gradient torque is negligible. For instance, Figure \ref{fig:radialsepN} compares each torque to others, which are calculated by adding the local value on each grid cell within a radius. As shown in Figure \ref{fig:radialsepN}, $N_\mathrm{m,prs}+N_\mathrm{grv}+N_\mathrm{prs}$ is negligible at most radii, except that around $10^{15}{\rm\,cm}$ $N_\mathrm{grv}$ is relatively important.
This is not inconsistent
with neglecting the magnetic pressure term in \citet{LKS2013}, since the authors analyzed torques integrated over a sphere rather than a grid cell. 

Figure \ref{fig:domTerm} shows the dominant torques in midplane within $100\au$ of model G, H, D, and A. To examine the forces with physical meaning, only the z-component of torques due to Lorentz force, thermal pressure gradient and gravity, are compared. 
Model I has similar results of dominant terms to model H and here for simplicity model I is not demonstrated. 
In model G, the gravitational torque rather than the magnetic torque dominates at late stage. It is consistent with the decline in the effect of magnetic field in this region. 
Gravity can play its special role within the inner parts, as we will further illustrate in Section \ref{sec:gi}.
In model H where the mass-to-flux ratio is smaller, however, the magnetic torque dominates in turn. It implies a stronger effect of magnetic torque on maintaining the angular momentum. 
In model D, the spiral pattern manifests where the pressure gradient torque is prominent. As mentioned previously, \citet{LKS2013} focus the analysis on the global torques so that the gravitational torque and pressure gradient torque are neglected. According to the dominant terms, however, local gravitational torque can be prominent. 

\begin{figure*}[htb]
	\centering
	\includegraphics[width=0.8\textwidth,trim={0 1cm 0 1.5cm},clip]{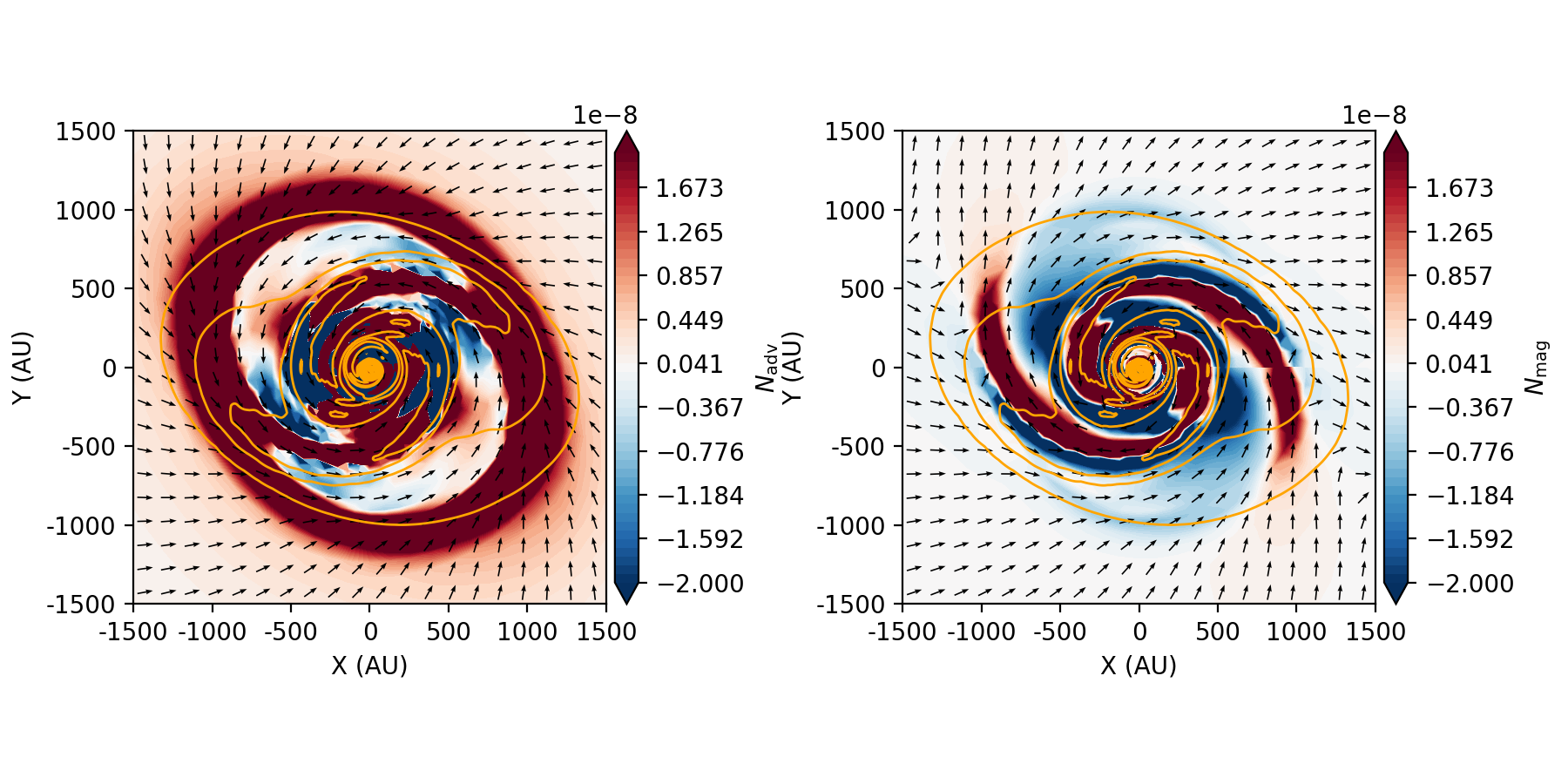}
	\includegraphics[width=0.8\textwidth,trim={0 1cm 0 1.5cm},clip]{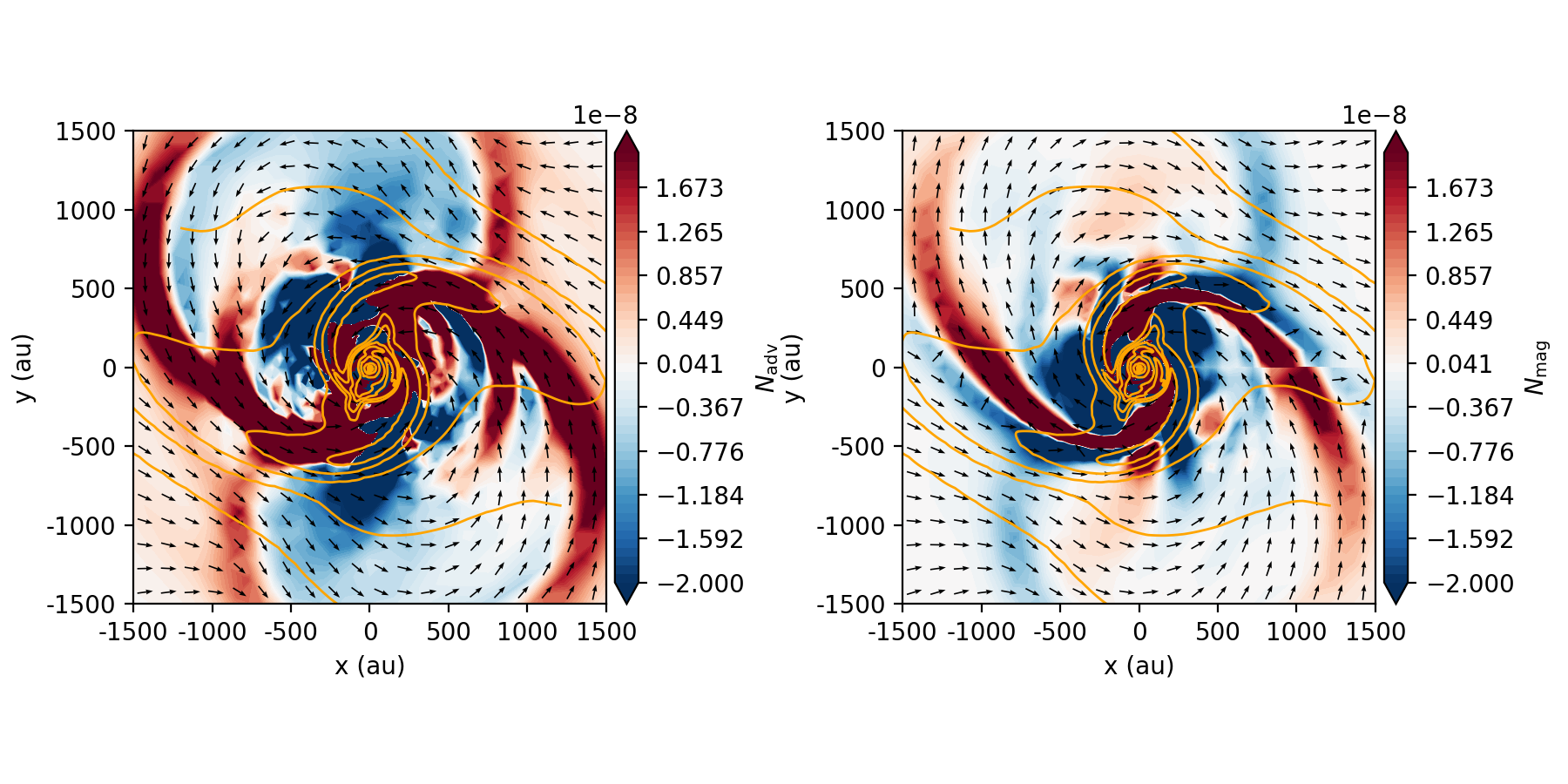}
	\includegraphics[width=0.8\textwidth,trim={0 1cm 0 1.5cm},clip]{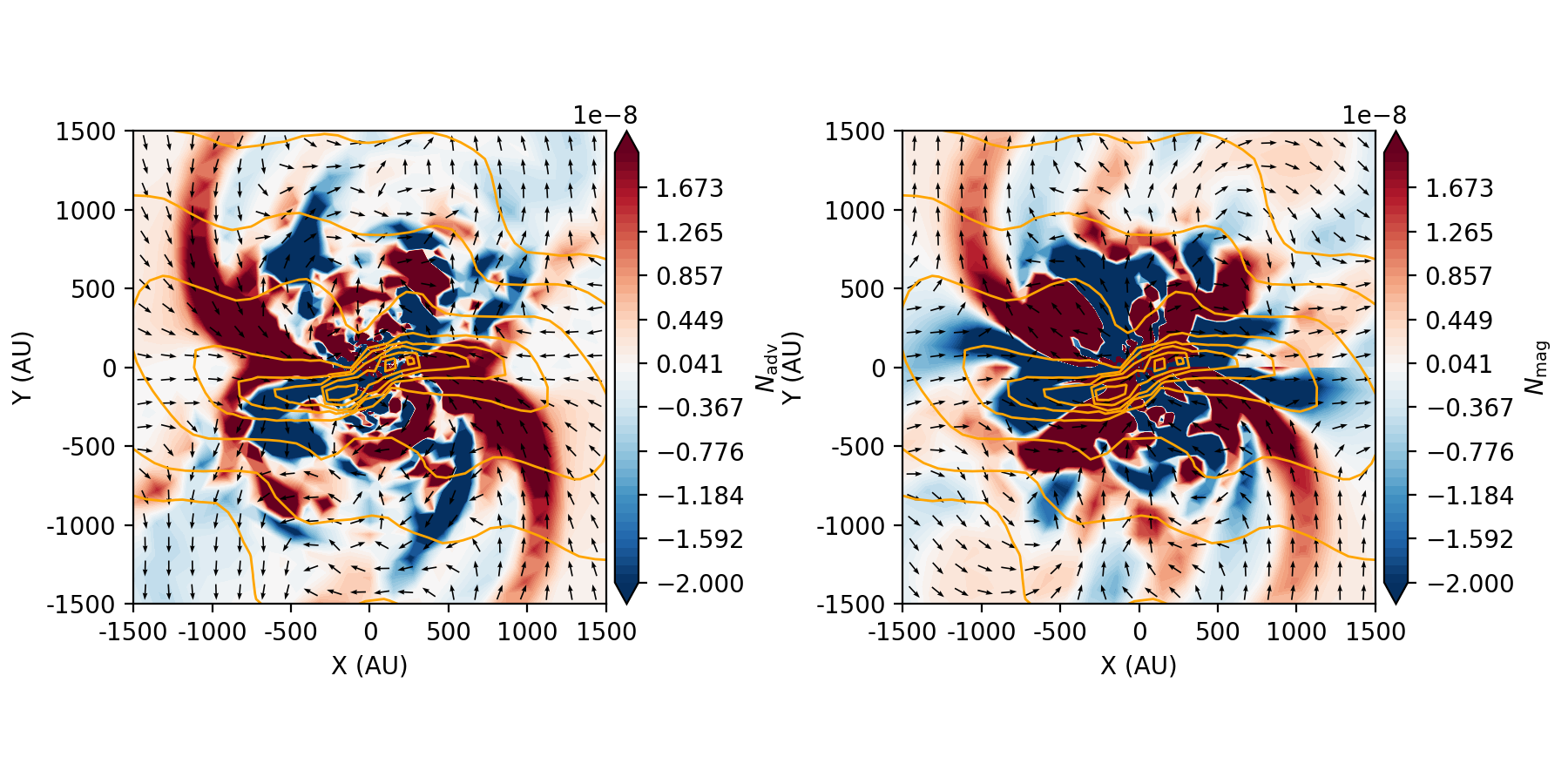}
	\includegraphics[width=0.8\textwidth,trim={0 1cm 0 1.5cm},clip]{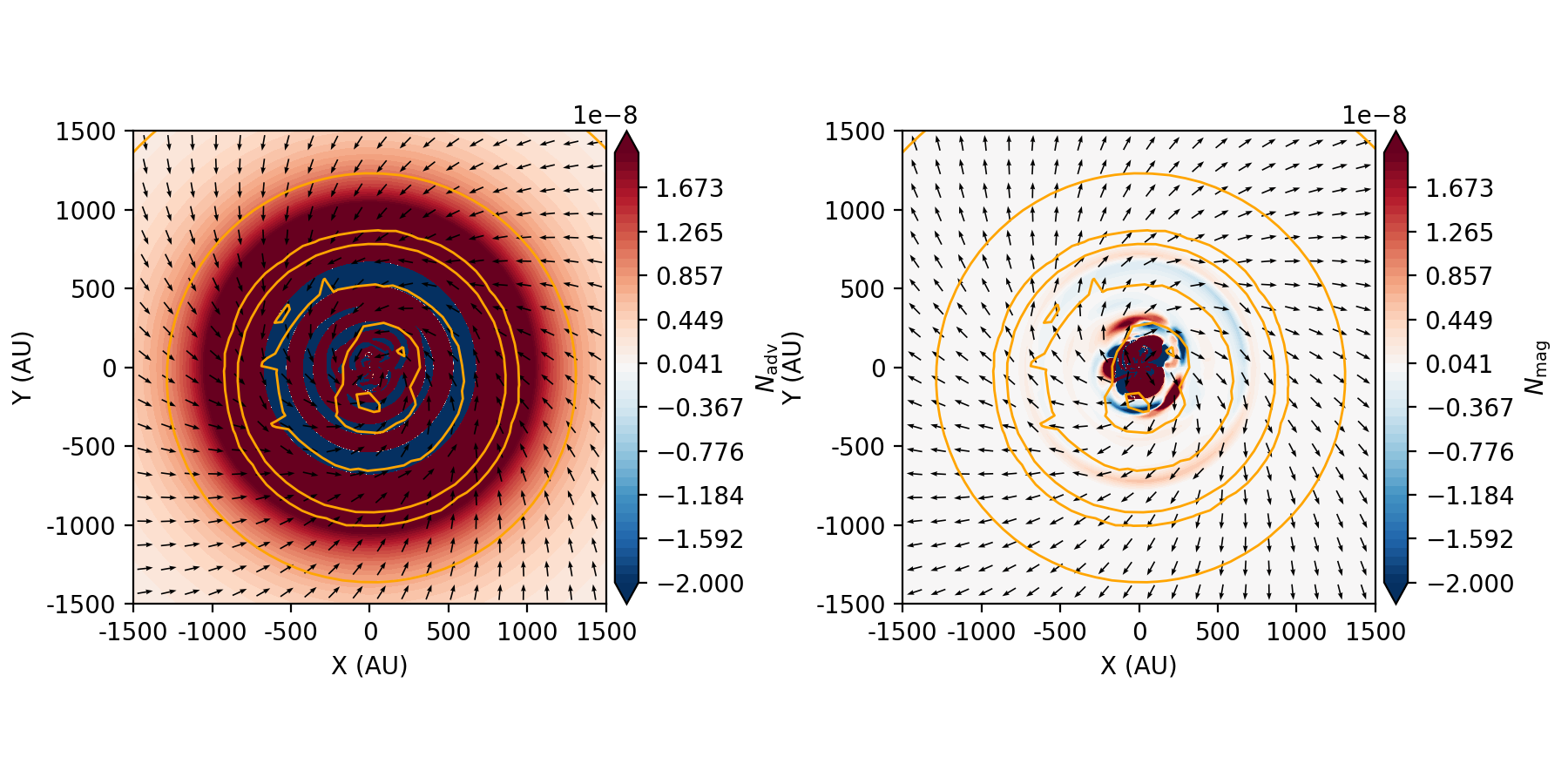}
	\caption{Color map of advective torque (left panels) overlaid by velocity (arrows) and column density (yellow contours), and magnetic torque (right panels) overplotted by magnetic field (arrows) and column density (yellow contours) on midplane of model G at frame 429; model H at frame 456; model I at frame 525; model A at frame 400.}\label{fig:ovplt}
\end{figure*}

Figure \ref{fig:ovplt} demonstrates the connection among torques, column density, magnetic field and velocity of models G, H, I, and A, respectively. 
The advective torque has no strict correspondence with velocity but the convergence of velocity. The latter usually relates to shock waves. Utilizing the method mentioned in Section \ref{sec:mthd_shock}, we detect shocks to fathom whether the ring-like and spiral structures are formed due to shock waves. 
The results will be shown in Section \ref{sec:shocks}.
The torques due to thermal pressure gradient and gravity are less essential to the ring-like expanding wave surrounding the spirals.
The spiral structure of advective torque follows the column density spirals. 
The top-right panel of Figure \ref{fig:ovplt} shows that the edge of positive and negative magnetic torque of model G strictly corresponds to the region where the direction of magnetic field reverses. For model H and model I, though the spatial distribution of magnetic torque looks chaotic, we can always find a pair of symmetric spirals where the magnetic field reverses and the magnetic torque changes its sign across the spirals. Yet model A shows smaller influential region of magnetic torque.

\subsection{Role of gravitational instability}\label{sec:gi}

Since a robust RSD is present in model G, the gravitational instability of the disk can be analyzed for this model, involving the formation of spiral structure. Disk mass and its induced potential well are essential factors of gravitational instability. Thus the disk-to-star mass ratio $M_{\rm d}/M_*$ can be utilized to reflect the underlying physics. For simplicity, $M_{\rm d}$ is approximated to the enclosed mass within $100\au$, at which the outer edge of the disk for model G is roughly located. $M_*$ is regarded as the central mass.

\begin{figure}[htb]
	\centering
	\includegraphics[width=0.5\columnwidth]{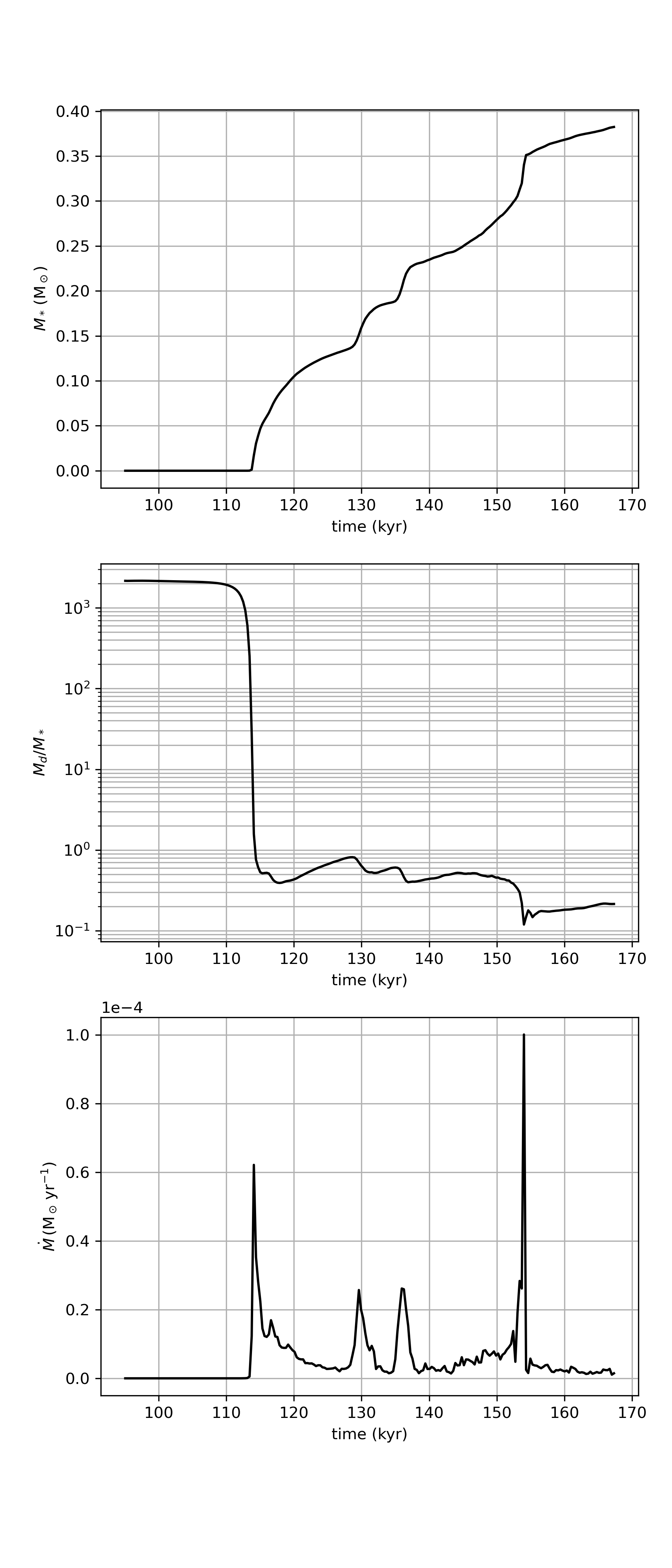}
	\caption{Central mass growth (top), disk-to-star mass ratio (middle), and mass accretion rate (bottom) of model G.}\label{fig:Gcm_d2s_mdot}
\end{figure}

Before efficient accretion begins (about $114.1\kyr$), the disk-to-star mass ratio is much greater than unity (see Figure \ref{fig:Gcm_d2s_mdot}). It rapidly declines below unity afterwards, when the disk starts to form and its mass accumulates. During the process the ratio decreases slowly, with some temporary increases in the ratio. The mass accretion rate behaves correspondingly. At around $129.9\kyr$, $136.3\kyr$ and $154.3\kyr$, another three bursts of rapid accretion occur, after an initial buildup of disk mass. They are most likely triggered by a sort of instability in the disk, as disk systems with large mass have tendency to be gravitationally unstable. 
Therefore, we calculate Toomre $Q$ parameter of model G to investigate the possible presence of gravitational instability.

\begin{figure}[htb]
	\centering
	\includegraphics[width=\columnwidth]{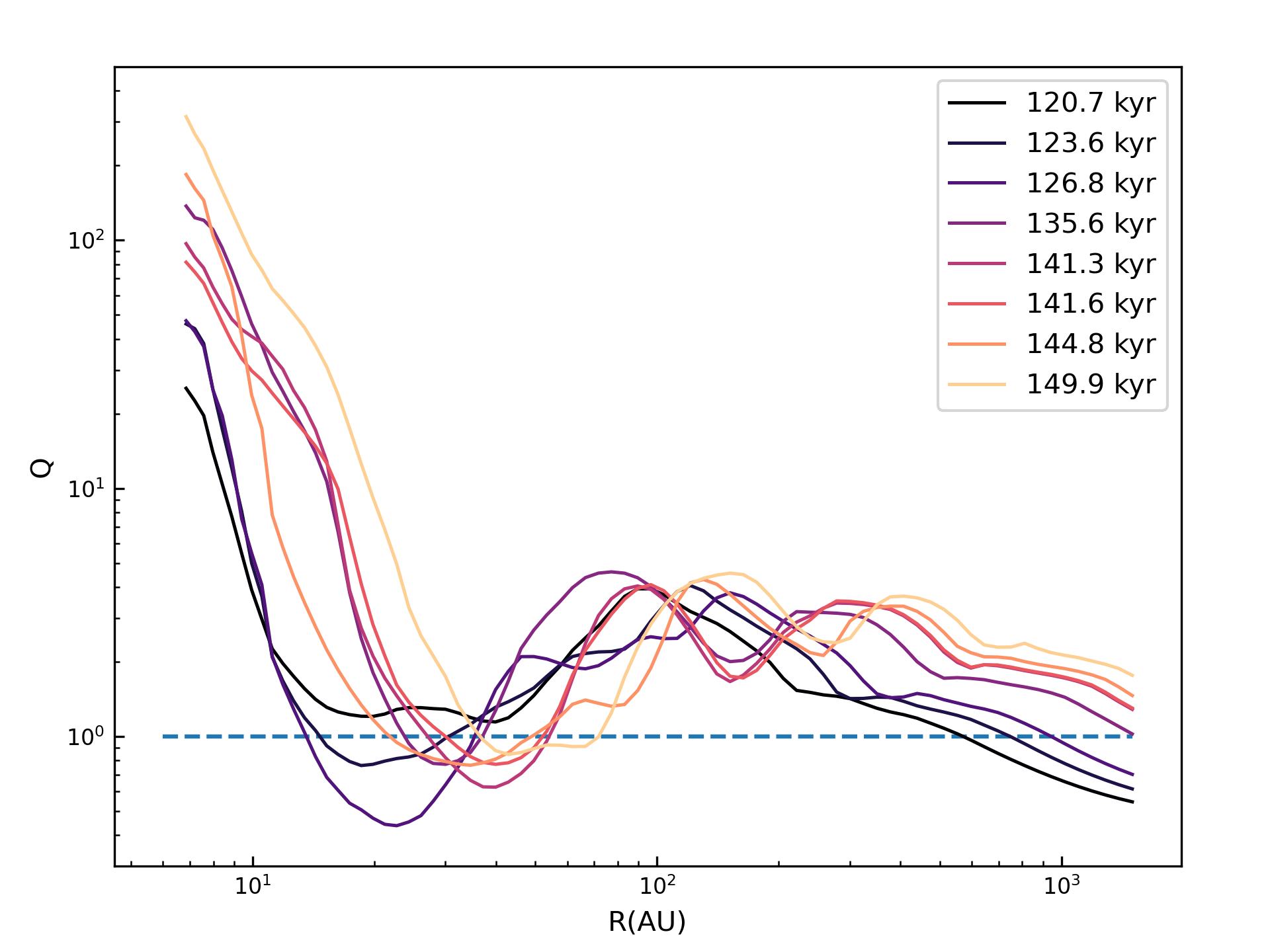}
	\caption{Integrated Q parameter in model G at frames between the first and last burst of rapid accretion. $R$ is the cylindrical radius on the disk plane.}\label{fig:GToomreQ}
\end{figure}

Figure \ref{fig:GToomreQ} shows the radial distribution of Toomre $Q$ of frames at which the disk-to-star mass ratio is relatively large. During this period, Toomre $Q < 1$ within a few tens of AU, which suggests that the gravitational instability grows in the disk. It may account for the disk wobbling \citep[see also Figure 7 in][]{MV2019} around $154.3\kyr$. The sharp decrease in disk-to-star mass ratio after $154.3\kyr$ agrees with the disk wobbling. At large radii, $1000\au$ for instance, though the gas is sparse, it flows inwards with little rotation. Then it leads to a relatively low value of $Q$.

\begin{figure}[htb]
	\centering
	\includegraphics[width=0.5\columnwidth]{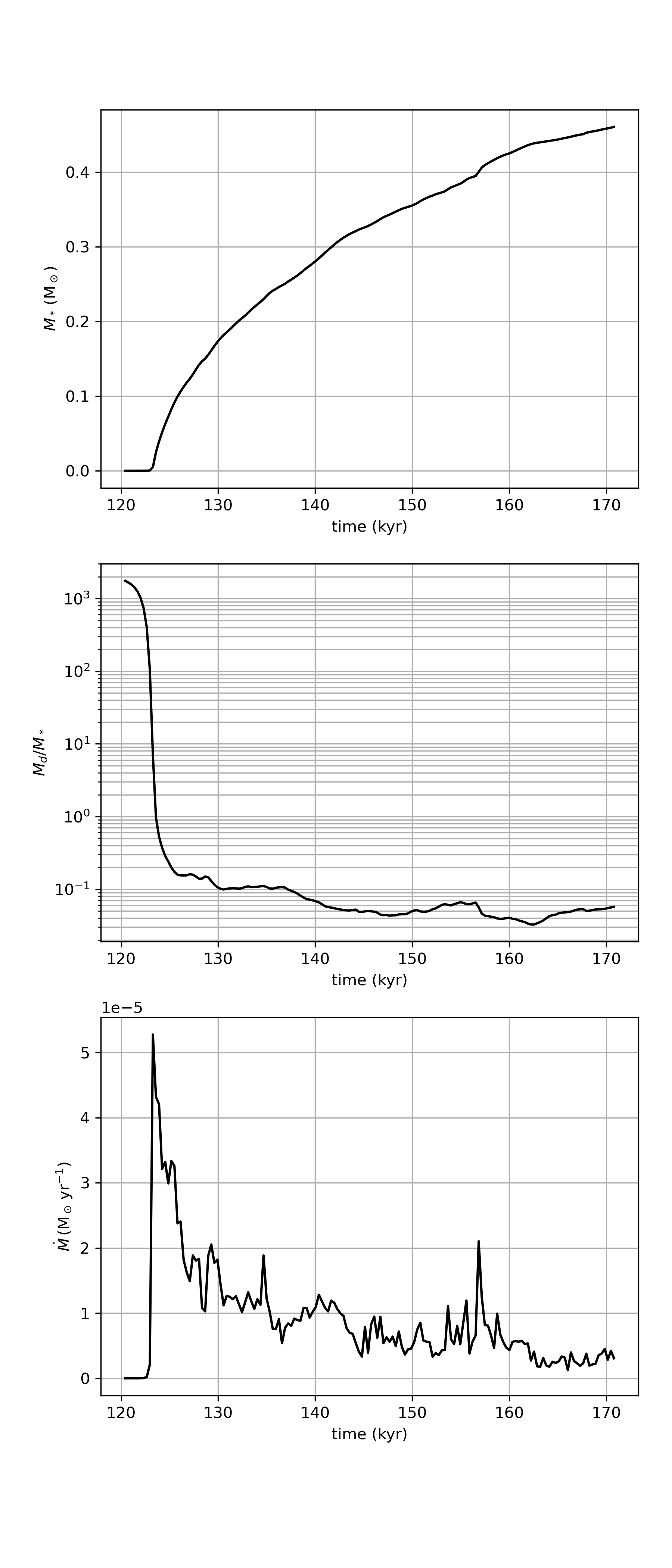}
	\caption{Same as Figure \ref{fig:Gcm_d2s_mdot} but for model H.}\label{fig:Hcm_d2s_mdot}
\end{figure}

\begin{figure}[htb]
	\centering
	\includegraphics[width=\columnwidth]{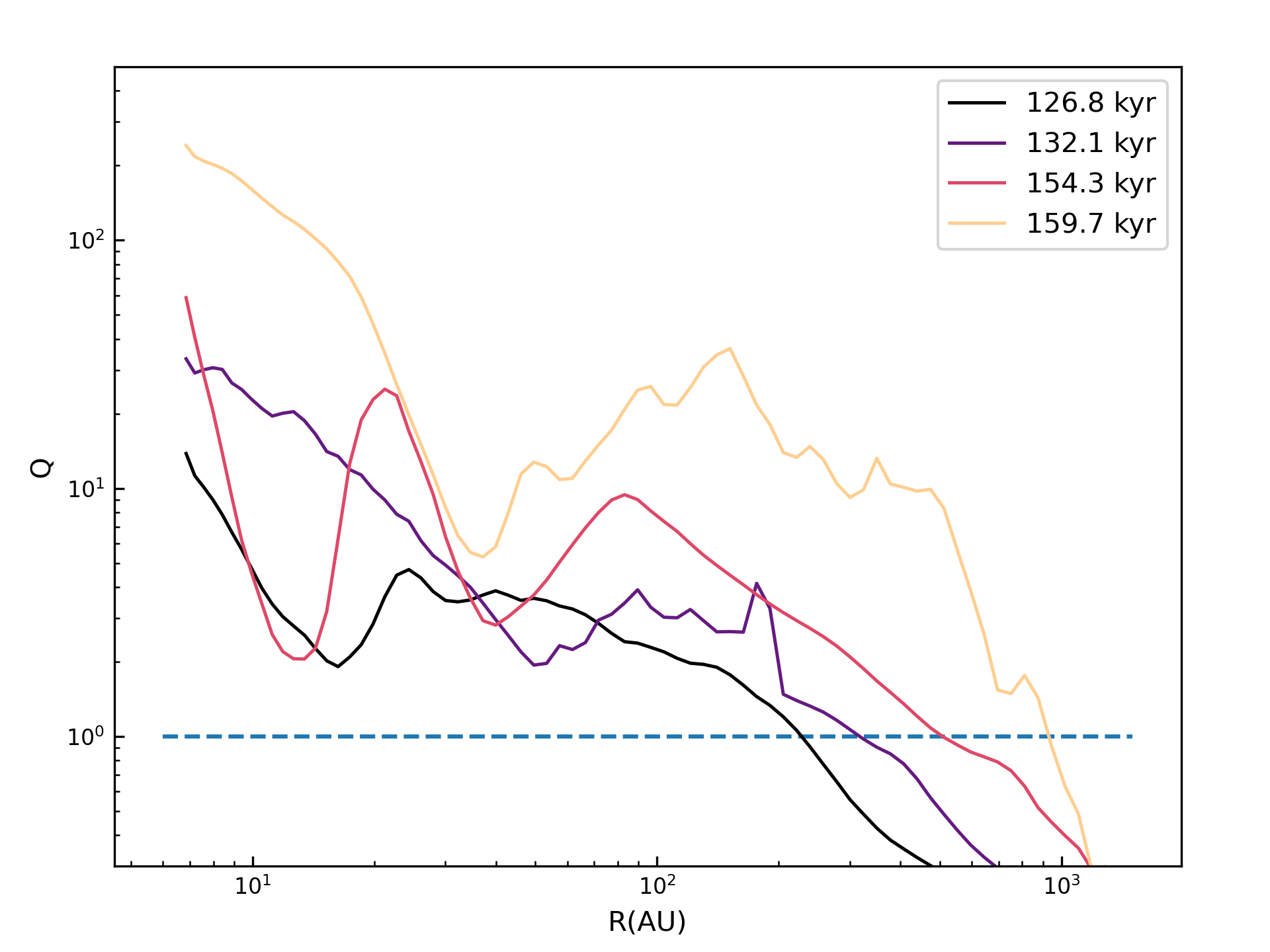}
	\caption{Integrated Q parameter in model H at selected frames when an inner disk is clearly distinguished. Thus the column density is derived along disk rotating axis. Same with Figure \ref{fig:GToomreQ}, $R$ is the cylindrical radius on the disk plane.}\label{fig:HToomreQ}
\end{figure}

Disk also forms in model H, even though it is porous as described in \citet{LKS2013} and \citet{MV2019} and disk size is clearly smaller than that in model G. Figure \ref{fig:Hcm_d2s_mdot} shows $M_{\rm d}/M_*$ is relatively small and new accretion substantially slows down after about $126\kyr$. Then we do the same analysis on model H\footnote{Here notice that in model H the disk precesses rather than rotating along the fixed z-axis. The column density needed for calculating Toomre $Q$ parameter should be integrated along the rotation axis. So the disk plane should be recognized. See Appendix \ref{sec:Append_diskplane} for the method used to determine the disk plane.} as model G. The Toomre $Q$ parameter exceeds unity most of the time (see Figure \ref{fig:HToomreQ}). Meanwhile, the gravitational torque does not dominate other terms (see Figure \ref{fig:domTerm}). It indicates the disk is gravitational stable in model H with stronger magnetic field than in model G. It implies that strong magnetic field suppresses gravitational instability, which is also suggested in \citet{LGCA2010}.

\begin{figure*}[htb]
	\centering
	\includegraphics[width=\textwidth]{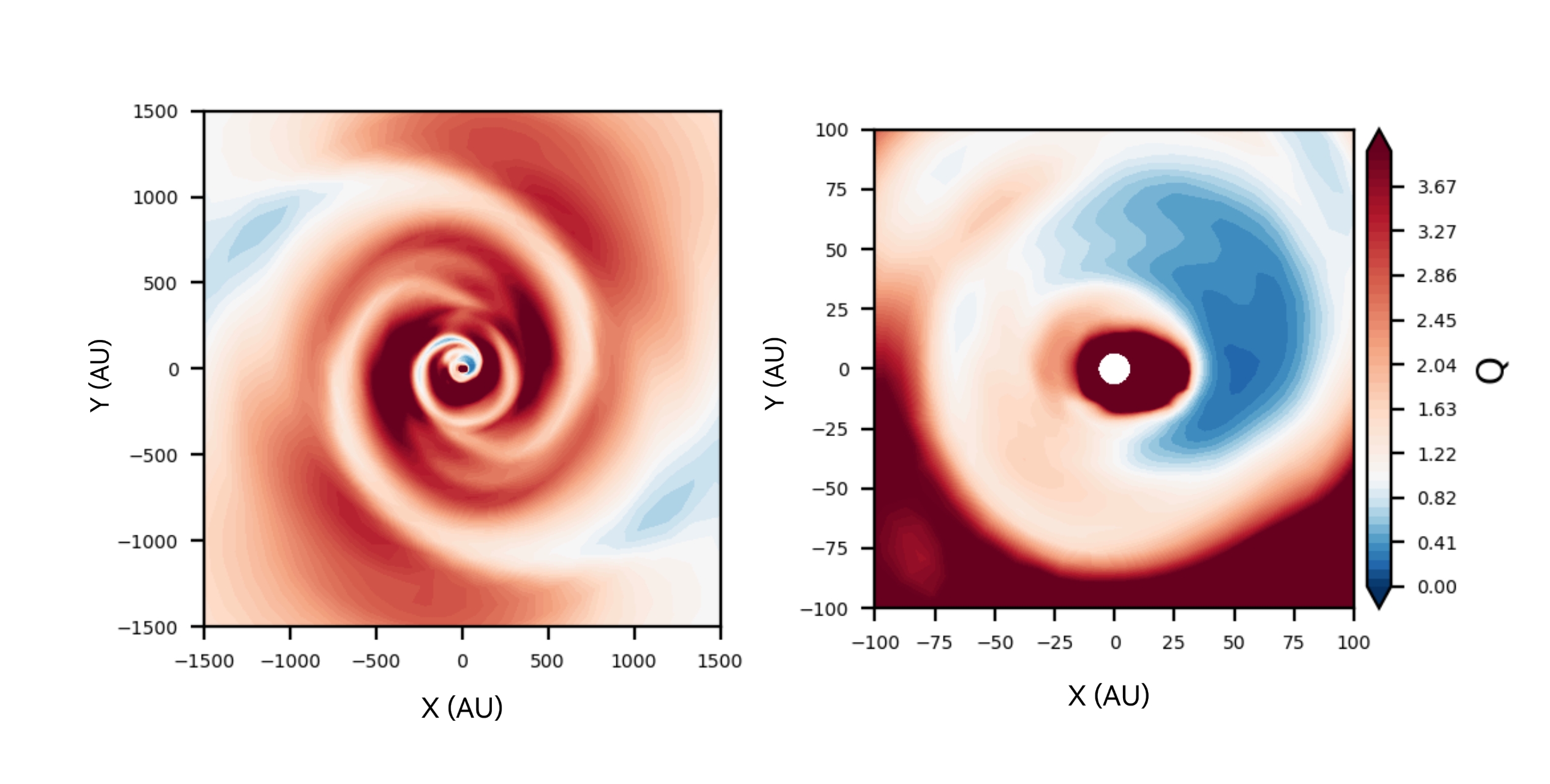}
	\caption{Toomre Q at midplane of model G at $149.9\kyr$. The right panel is a zoom-in of the panel of the left. White color represents $Q=1$. }\label{fig:2dGToomreQ}
\end{figure*}

\begin{figure*}[htb]
	\centering
	\includegraphics[width=\textwidth]{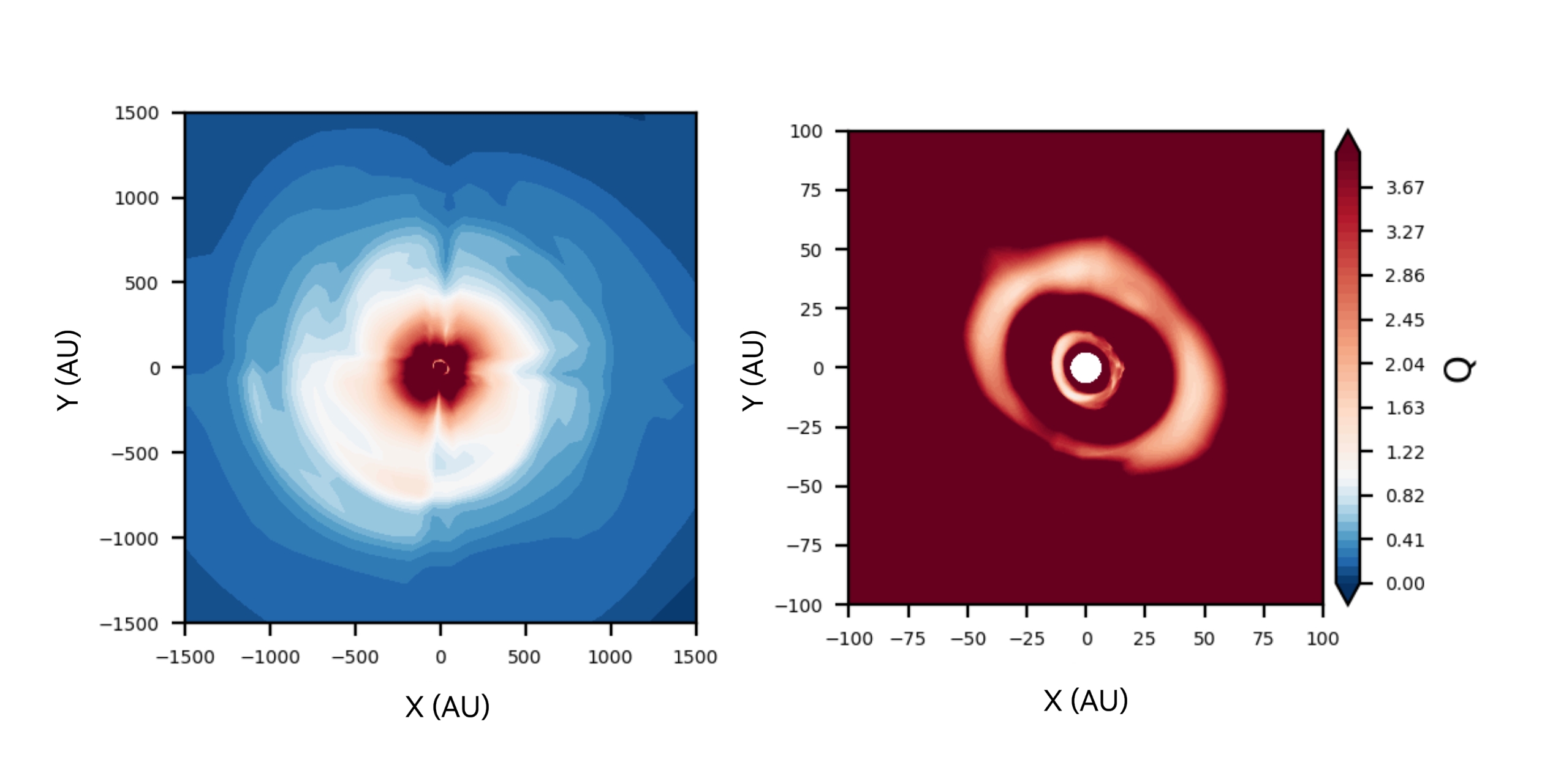}
	\caption{Same as Figure \ref{fig:2dGToomreQ} but at disk plane for model H at $154.3\kyr$. }\label{fig:2dHToomreQ}
\end{figure*}

The 2D figure at disk plane illustrates a more clear relation between the spiral structure and Toomre Q. As shown in Figure \ref{fig:2dGToomreQ} of model G, the spiral structure in column density correlates with the relatively low value region of Toomre Q. In contrast to the model G, no apparent spiral structure in Toomre Q (see Figure \ref{fig:2dHToomreQ}) of model H is identified within $100\au$. The central disk of model H, whose size is much less than $100\au$, is not affected by gravitational instability.

\subsection{Magnetic spirals}\label{sec:spirals}

\begin{figure}[htb]
	\centering
	\includegraphics[width=\columnwidth]{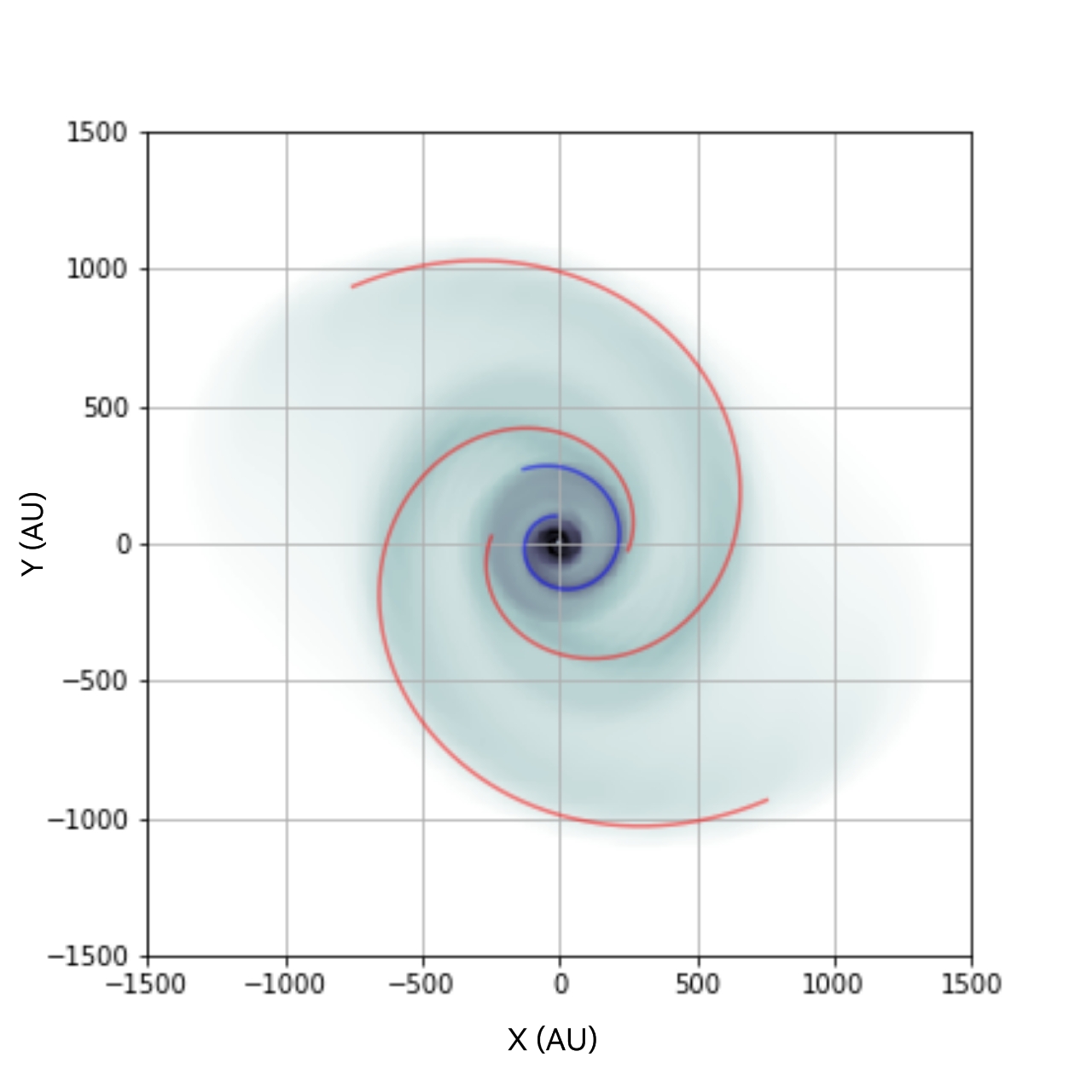}
	\caption{Density overlaid by fitted spirals. The one-arm spiral in the inner region (blue curve), and the two-armed spiral in a more outer region (red curve), are fitted as shown in Section \ref{sec:spirals} using the method of Section \ref{sec:mthd_spiral}.
	}\label{fig:spiralfit}
\end{figure}

According to torque distribution at midplane (Figure \ref{fig:ovplt}) and column density (Figure \ref{fig:spiralfit}, for more examples see \citealp{MV2019}), misalignment models (G,H,I,D) show large-scale stable two-armed spirals. As shown by \citet{LKS2013}, magnetic field is wrapped around like a snail shell, and the spiral arms are located between two sheets of magnetic fields with opposite signs, where the magnetic field reverses. 
As magnetic torque and magnetic field of misaligned models show in Figure \ref{fig:ovplt}, the large-scale spirals are always followed by the reversed magnetic field which is the consequence of misalignment and rotation. 

While in principle the magnetic spirals are simple results of geometry, in practice, we have to separate outer spirals and inner spirals. The outer spirals remain in principle a coherent structure, whereas the inner spirals are prone to chaotic effects. In model H, we get rings and gaps, further elaborated in Section \ref{sec:warped}. In model G, other forces come to take part.

As a special case, in model G a small-scale one-armed spiral (Figure \ref{fig:spiralfit}) is identified, which corresponds to the inner disk. We show in Section \ref{sec:gi} gravitational instability can grow in such magnetized structure.
The one-armed spiral corresponds to the mode $m=1$ for hydrodynamic density wave. 
If we estimate the geometrical properties of the spirals, the logarithmic function fits those spirals well with slight deviation. The fitting functions have parameter of $b=6.0$ for the one-armed spiral structure and $b=3.5$ for the two-armed spiral structure. The corresponding pitch angles are about $9\degree$ for the inner spiral and $16\degree$ for the outer (See Figure \ref{fig:spiralfit}).

It should be noted that gravitational instability and magnetic spirals are not mutually exclusive effects. Both phenomena can be nonlinearly coupled. Even when gravitational instability becomes significant, system is still under considerable magnetic forces. A potential example of this coupling could be that launching of a blob in the model G. There as a combination of gravitational instability and mass accretion, a blob of gas is launched outwards from near the system centre. Its launch is likely slowed down by the toroidal magnetic field in the system, making the blob fall back down within an elliptical orbit. Without the magnetic effect, blob could be expected to propagate farther away - perhaps becoming a genuine fragment.

\subsection{Central mass growth and angular momentum redistribution}\label{sec:growth}

In Model G, four significant bursts of central mass growth occur (Figure \ref{fig:Gcm_d2s_mdot}). The first would be the beginning of the inside-out collapse. We choose the frames at these four peaks to investigate the torques and angular momentum in detail ($114.4\kyr$, $129.9\kyr$, $136.6\kyr$, $154.3\kyr$), and those who are $3.169\kyr$ before the peaks ($111.2\kyr$, $126.8\kyr$, $133.4\kyr$, $151.2\kyr$) at which times the mass accretion rate is relatively low. 

\begin{figure*}[htb]
	\centering
	\includegraphics[width=\textwidth]{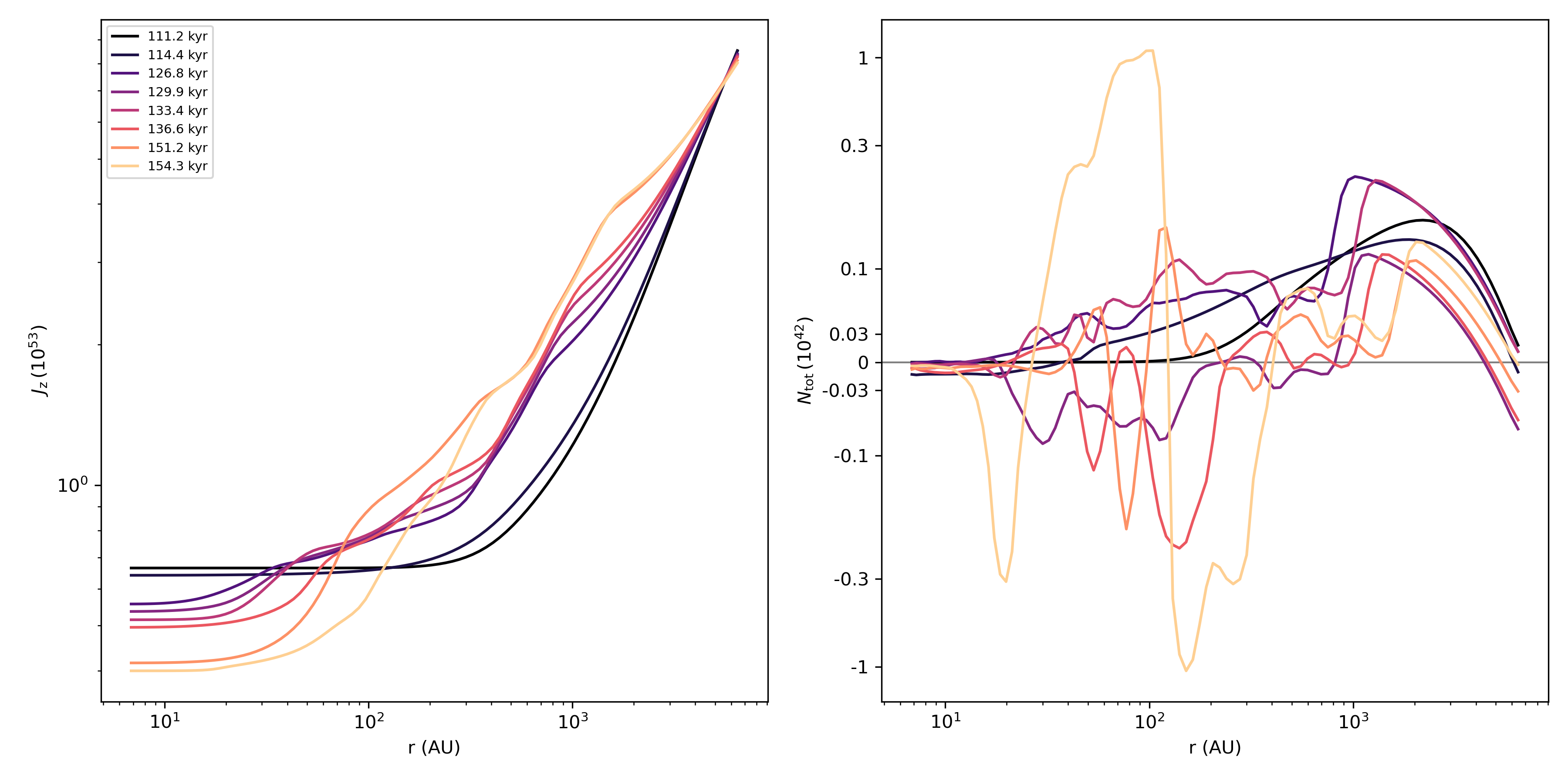}
	\caption{Z-component of angular momentum (left) and total torque (right) within spheres of a given radius $r$. Different colors and line styles depict quantities at different time-frames. The angular momentum is plotted in units of ${\rm g\,cm^2\,\second^{-1}}$ and torques in ${\rm dyn\, cm}$. The left panel is plotted in logarithmic scale and the right in symmetrically logarithmic scale with linear scale within $\pm0.2$.}\label{fig:radialN}
\end{figure*}

As presented in Figure \ref{fig:radialN}, the z-component of angular momentum $J_{\rm z}$ redistributes during evolution, while the total amount marginally conserves. In the inner tens of $\au$, $J_{\rm z}$ decreases with time, while in the outer thousands of $\au$, the angular momentum rises with time. The enclosed angular momentum of a sphere within around $100\au$, for instance, increases first and then drops as the system evolves, especially the drop is significant at the fourth peak of mass accretion rate. It is accordant with the expectation of accretion process.

The total torque here $N_{\rm tot}$
equals the RHS of Equation \ref{eqn:jz}. In the inner $10\au$ the total torque is always negative and in the outer $1000\au$ it remains positive. The peak around $1000\au$ corresponds to the expanding wave shown in the color map of torques (Figure \ref{fig:ovplt}). The torque varies significantly in between. At frames of low mass accretion rate, the total torque tends to be positive at most of the radii. Meanwhile, except the first burst of accretion when the inside-out collapse just begins, the total torque can be negative in the scale of $100\au$. It is evident that the burst of accretion accompanies tremendous redistribution of angular momentum.

\begin{figure*}[htb]
	\centering
	\includegraphics[width=\textwidth]{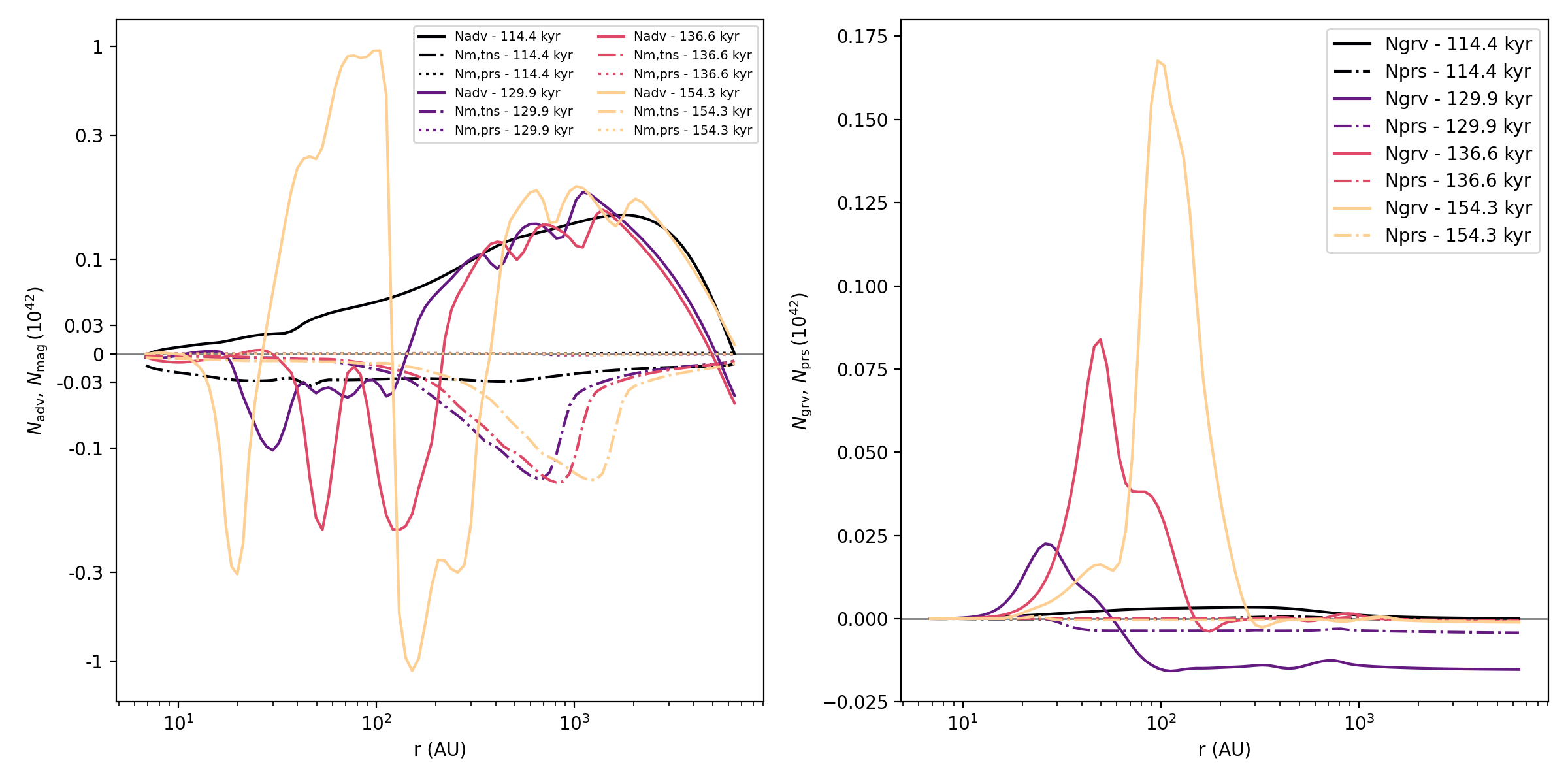}
	\caption{Advective torque (solid lines), magnetic torque (dot-dashed lines), gravitational torque (solid lines), and pressure gradient torque (dot-dashed lines). The left panel is displayed in symmetrically logarithmic scale with linear scale within $\pm0.2$ and the right in linear scale. Torques given in ${\rm dyn\, cm}$.}\label{fig:radialsepN}
\end{figure*}

In order to investigate the mechanisms in detail, we plot each torque at these four peaks of mass accretion rate in Figure \ref{fig:radialsepN}. The torques due to magnetic pressure and thermal pressure gradient are zero or negligible due to insignificant numerical deviation as expected. In addition, we notice that the gravitational torque around $100\au$ can make minor contributions to angular momentum transfer, even though it is nearly one order of magnitude smaller than the advective torque. Moreover, at late stage the gravitational torque exceeds the magnetic torque, which is also implied in Figure \ref{fig:domTerm}. The trend of its increasing with time suggests that the gravitational torque potentially becomes comparable to the advective torque under favorable conditions.

\subsection{Locating shocks}\label{sec:shocks}

\begin{figure}[htb]
	\centering
	\includegraphics[width=\columnwidth]{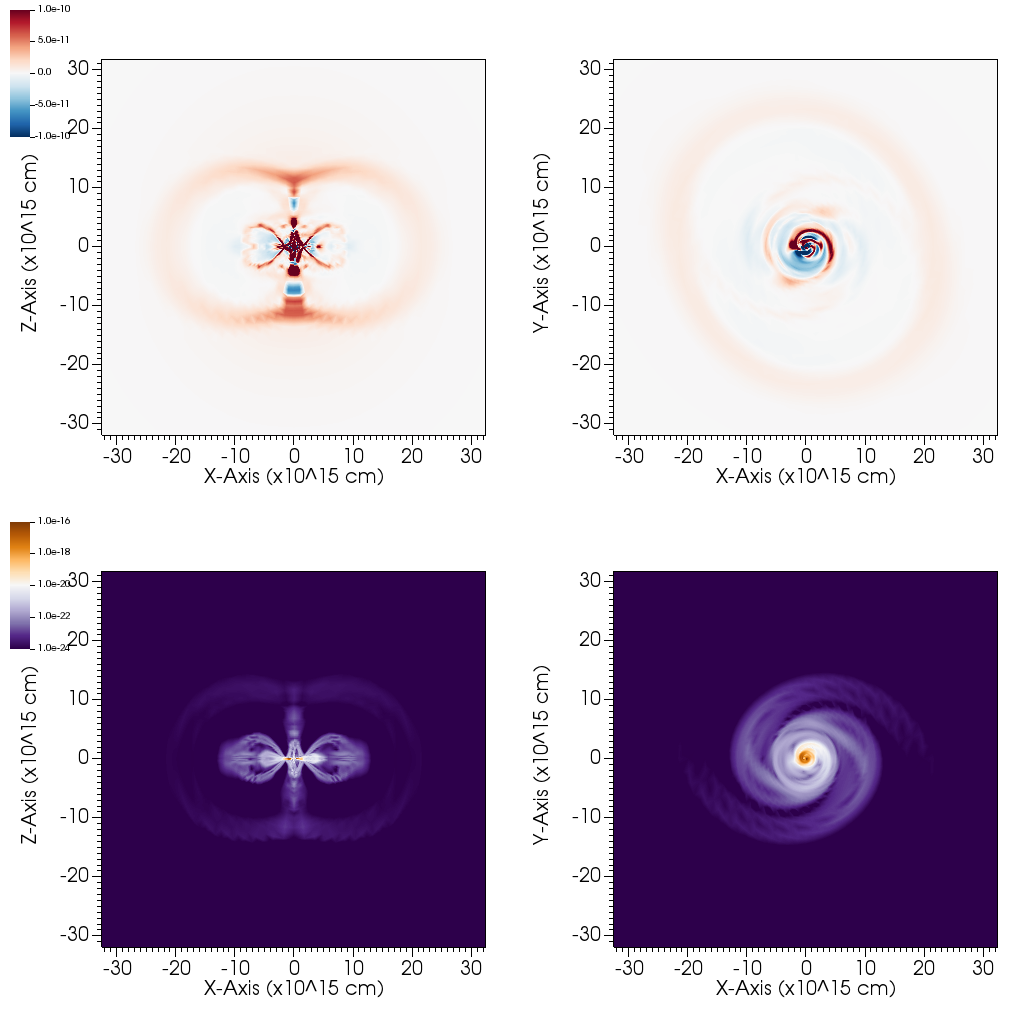}
	\caption{Cut of convergence of velocity (upper panels) and thermal pressure gradient (lower panels) at x-z plane (left column) and x-y plane (right column) of model G at $144.5\kyr$. White color presents $-\nabla\cdot\mathbf{v}=0$ (upper panels) and $|\nabla p|=10^{-20}\rm{g\,cm^{-2}\,s^{-2}}$ (lower panels).}\label{fig:G_divv_gradp}
\end{figure}

In the Figure \ref{fig:G_divv_gradp} shown at $144.5\kyr$,
the noticeable pattern of convergence of velocity and thermal pressure gradient is an X-shape structure on the x-z plane and a spiral structure on the x-y plane. 
In addition, we see an expanding wave naturally appearing as a ring structure behaving as an inside-out collapse.

Viewed on the x-y plane, it shows clear inner one-armed spiral structure of the compression region. 
Besides, viewed on the x-z plane an x-shape structure exists for a long run.
The x-shape structure represents an inner edge compression layer of the ``pseudodisk'', as the arms wrap around themselves while spiralling inwards, and the outer ring is that expanding wave front of the inside out collapse. 
The complex structure along z-axis may be influenced by boundary conditions, which therefore cannot be fully trusted for conclusions.

\begin{figure}[htb]
	\centering
	\includegraphics[width=\columnwidth]{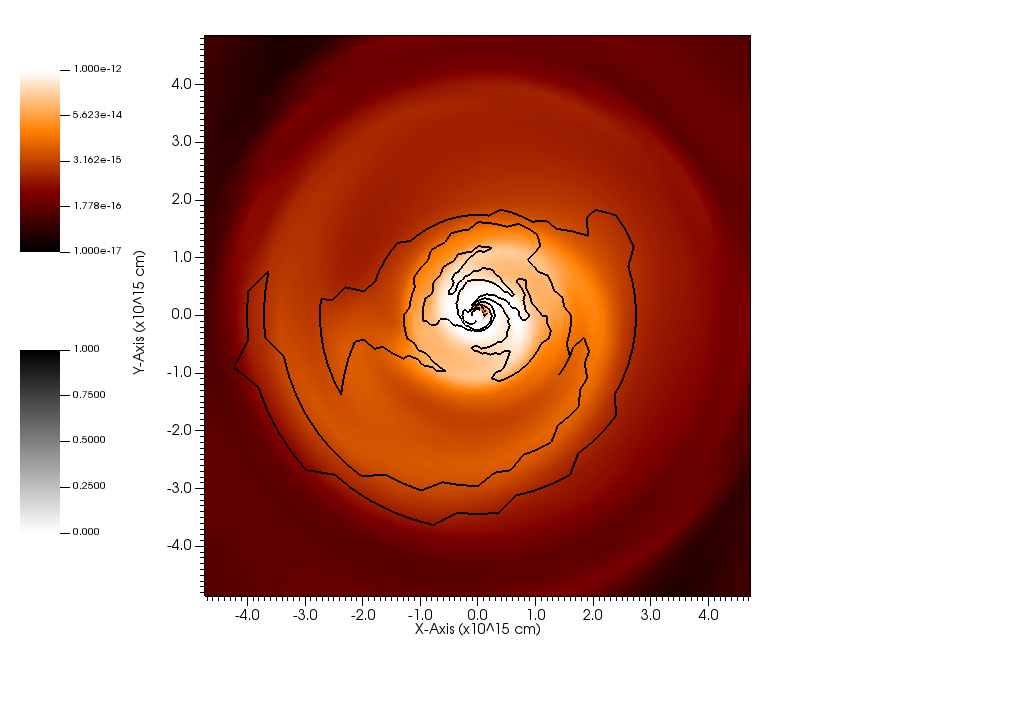}
	\caption{Density (colormap) overlaid with shock (region within contours) at midplane of model G. To locate shock, the threshold is set to $10^{-20}\rm{s^{-1}}$ for convergence of velocity and $10^{-21}\rm{g\,cm^{-2}\,s^{-2}}$ for thermal pressure gradient. The contours encircle the region that meets the conditions of both thresholds.}\label{fig:Gshock}
\end{figure}

Strongest candidate for shocks are seen in the inner disk, which are identified by the combination of conditions mentioned in Section \ref{sec:mthd_shock}. In Figure \ref{fig:Gshock}, the shock front is tightly related with the inner spiral. In this sense, we regard this spiral pattern in shock as the so-called spiral shock driven by the rotation of the system.
The x-shape pattern on the other hand does not fulfil equal conditions, and therefore the spiral remains as the most substantial shock.

We have in Figure \ref{fig:Gshock} the presence of a shock front which is associated both to the outer and the inner portions of the spiral structure of the disk model G. The shock portion located on the outer part of the spiral fulfills the expected role of an accretion shock (being the boundary between fast infalling material and a region of slower accretion with significant rotational support). This role is modified by the break of axisymmetry induced by the presence of spirals both in the accretion channels and in the disk structure, so it does not separate matter as simply a matter of radial location on the midplane, but mediated through the spiral structure. The shock is reported as clearly present, having passed the stringent two threshold criteria limits given in Figure \ref{fig:Gshock} (less stringent threshold criteria show wider regions as candidates for perhaps weaker shocks).
This spiral-associated shock region presents no singly-defined centrifugal or accretion shock radius, rather a range that depends on azimuth around the disk spiral structure. We are dealing here with a more smooth and gradual accretion than in the more traditional axisymmetric models, making a less clear boundary between the inner disk and its surroundings. Some of these transition features might represent sharp lines marking the limits of the outer parts of the disk spirals, and the inner parts of the accretion channel spirals. It is possible that the magnetized shock structures observed in the simulation are either C-shocks or J-shocks, potentially distinguishable in local or global simulations of very high resolution. Features with strong effects in the observability of shocks, such as heated gas, may be desired in future work along these lines of research.

\subsection{Warped disk and rings}\label{sec:warped}
\begin{figure*}[htb]
	\centering
    \plottwo{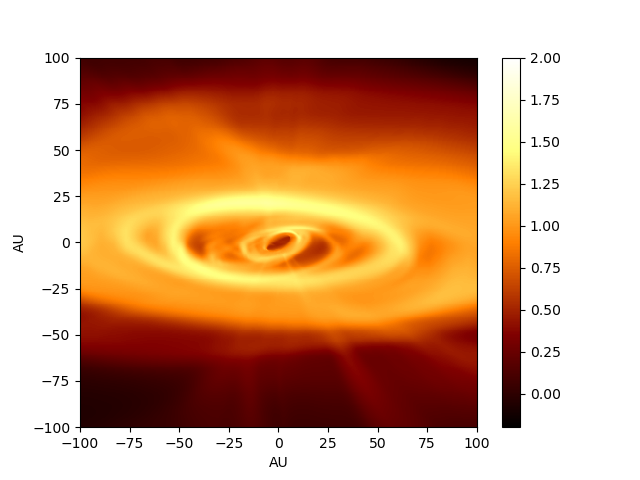}{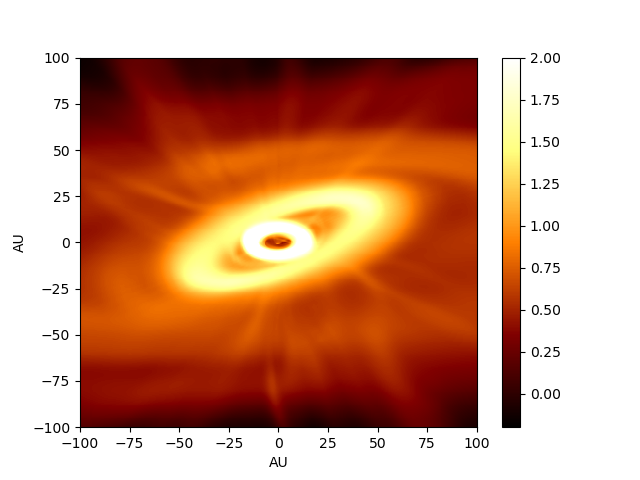}
	\caption{Column density of model H at $148.3\kyr$ (left panel) and at $154.3\kyr$ (right panel). The colormap displays the $\log_{10}$ of the column density in units of ${\rm\,g\,cm^{-2}}$.}\label{fig:Hcolden}
\end{figure*}

\begin{figure*}[htb]
	\centering
	\includegraphics[width=0.8\textwidth]{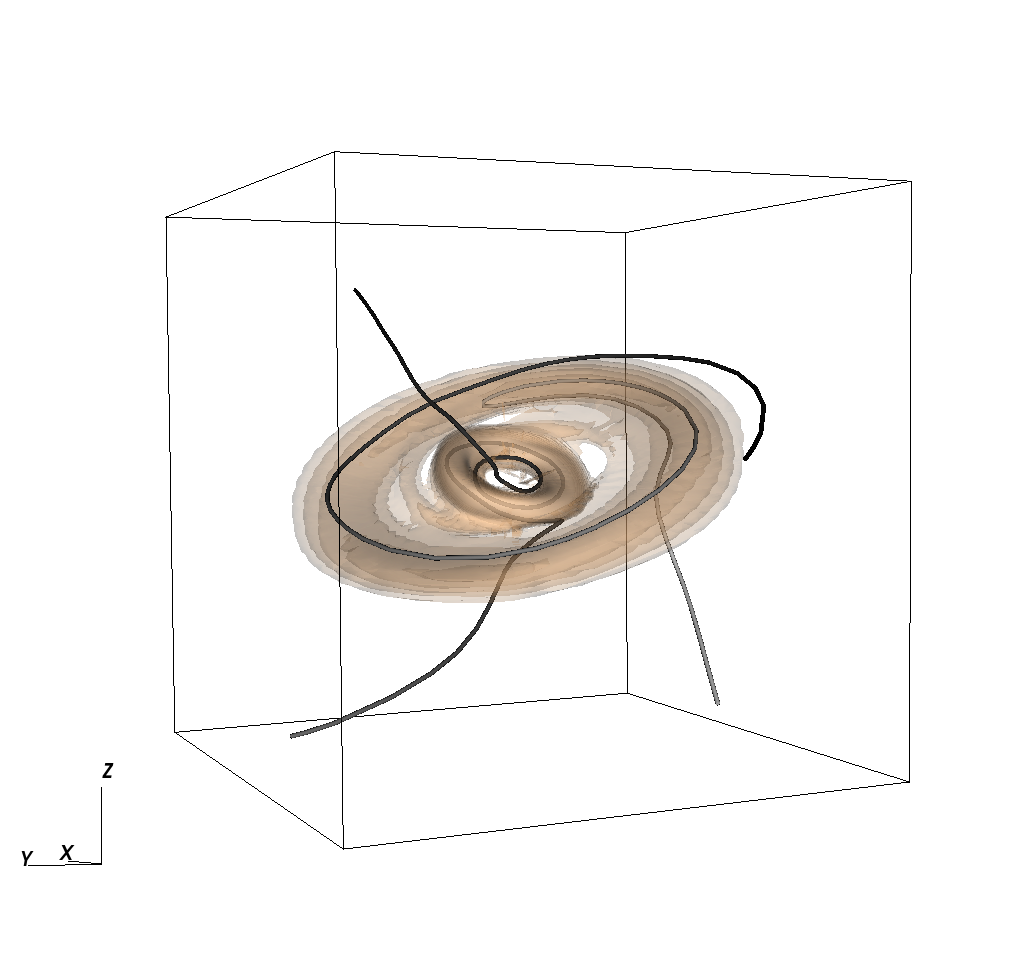}
    \caption{Density (iso-surfaces) with two embedded magnetic field lines (gray curves) whose start points are selected within the inner and outer disks for each. The box size is $133.69^3\au^3$ in Cartesian coordinate. The density iso-surfaces are logarithmically selected by 10 levels between $3.0
    \times10^{-14}{\rm\,g\,cm^{-3}}$ and $1.0
    \times10^{-12}{\rm\,g\,cm^{-3}}$.}\label{fig:Hrings}
\end{figure*}

\begin{figure*}[htb]
	\centering
	\includegraphics[width=\textwidth]{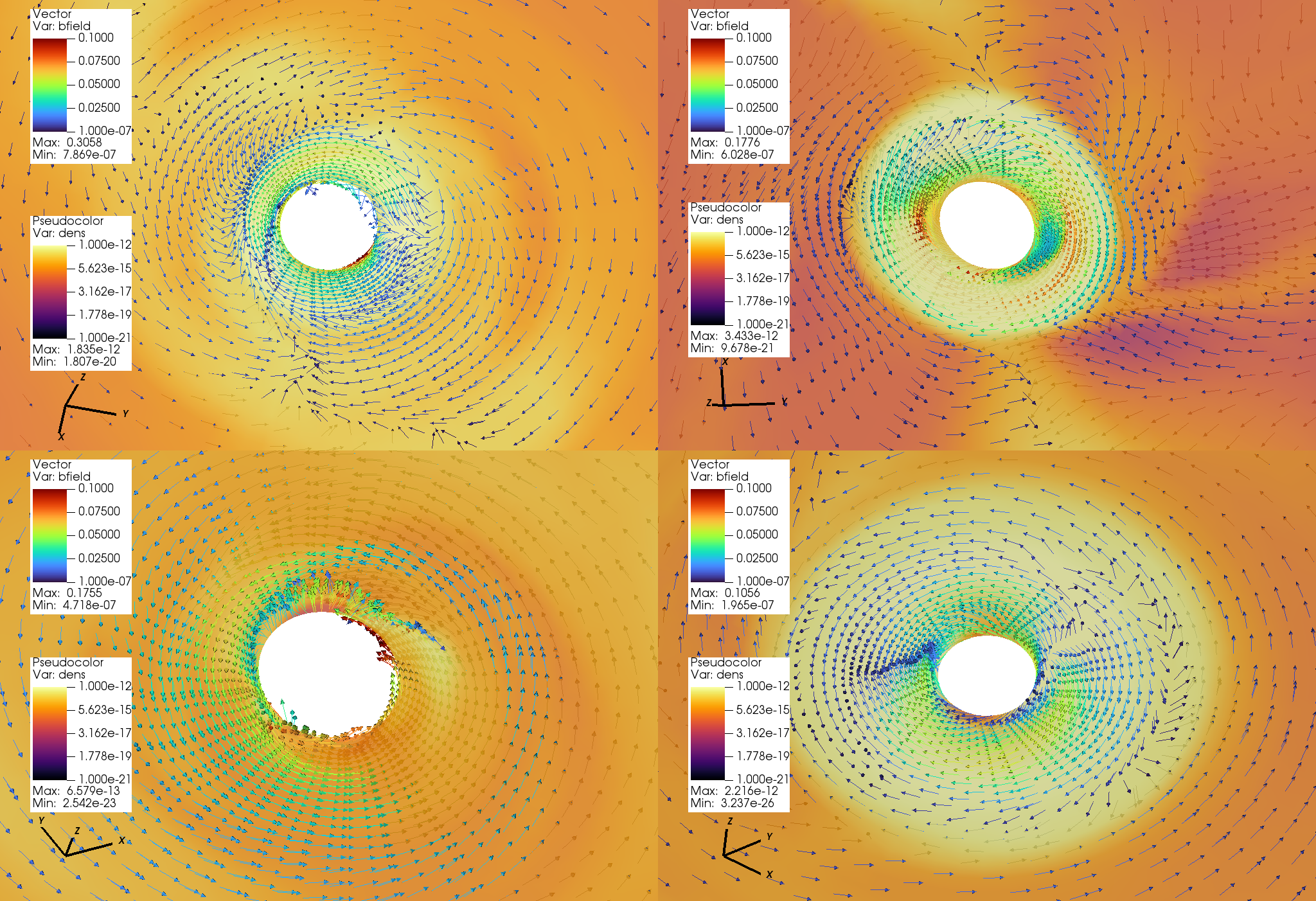}
	\caption{Density on the disk plane overlaid by magnetic field arrows for model H at different frames: $136.9\kyr$ (upper left), $154.3\kyr$ (upper right), $161.3\kyr$ (lower left), $167.6\kyr$ (lower right). Each panel has different viewing angle which is demonstrated by the triad rotation axis on the bottom-left corner of each panel.}\label{fig:Hfield}
\end{figure*}

In model H we notice the disk would evolve to substructures, warped disk and rings (see Figure \ref{fig:Hcolden}). They are not steady but evolve into each other back and forth. Disk is warped under the influence of precession, while outflow is driven along the z-direction. The inner and the outer rings have different inclination angles, and they have a gap of about the size of $10\au$ in between (also see Figure \ref{fig:Hcolden_cntr}).

The magnetic field lines interspersed in the rings show different paths (Figure \ref{fig:Hrings}). The inner substructure is more relevant to toroidal magnetic field; the outer is opposite. In Figure \ref{fig:Hfield} shown, when the inner disk is dense and compact (top left and bottom right panels), the embedded magnetic field reverses inside the disk; but when it is dispersed (top right and bottom left panels), the field direction does not reverse.

In model H, the disk is not rotating with a circular orbit \citep[Also suggested in][]{MV2019}. There is a tightly spiralling inward flow. 

Spirals are present in this flow, and through mechanisms of angular momentum change, the spiraling inflows largely circularize their orbit and form rings. For instance, at time $154.3\kyr$ two salient rings at radii of $20\au$ (inner ring) and $50\au$ (outer ring) are observed to feed from the inflowing spirals.

As the first row in Figure \ref{fig:h5} shows, ripples in the radial direction exist in the torques.
Such ripples in the torques result in the local increases in angular momentum at the radii of some rings, as well as decreases at their surroundings.  Migration of gas takes place according to these variations in the content of angular momentum at different radii, redistributing the gas.  Rings and gaps then form during this angular momentum redistribution.

When the inner disk is formed, the field is curved heavily. Then from the removal of angular momentum due to polar outflow driven by a possible magnetic tower jet results in both accretion and ejection of surrounding matter, the rings are dispersed. In such a case the magnetic field would play a dominant role in angular momentum transfer.

Loss of mass and angular momentum in outflows is then a possible cause of weakening and potentially destroying transient disk structures, as seen here in model H. While model D has a different structure, its transient disk is destroyed at least in part due to outflows.

The transient outflows observed in this set of simulations are to be distinguished from mature outflows such as those presented in \citet{Shang2020}; these mature outflows start from a long-duration magneto-centrifugal wind (such as the steady state flow in \citealp{SNOS1995}), and are largely governed by their interaction with the ambient medium. By contrast, the outflows presented here are transient structures of various origins, connected to the processes of disk formation; they have however in common that both kinds of outflows are channels for mass loss and angular momentum transport.

\subsection{Dependence on ambipolar diffusion}\label{sec:AD}

With regards to the existing dataset, our focus has been in ideal MHD scenario, and non-ideal MHD effects have been excluded. However, previous literature suggests that non-ideal MHD effects are crucial for disk formation during core collapse at least in the first core phase \citep[e.g.][]{Duffin2008,Mellon2009,Tsukamoto2015,Marchand2016,Masson2016,Hennebelle2016,Wurster2016,Zhao2016,Vaytet2018}. To address the influence of non-ideal MHD effects we examine preliminary results in this point by considering ambipolar diffusion as a case of weak ionization. The formula of drag force per unit volume exerted on the neutrals by the ions reads 
\begin{equation}
    \mathbf{f}_{\rm d}=\gamma\rho_{\rm n}\rho_{\rm i}\mathbf{v}_{\rm d},
\end{equation}
where $\gamma=\frac{\left<\omega\sigma_{\rm in}\right>}{m_{\rm n}+m_{\rm i}}$ is the drag coefficient with $\left<\omega\sigma_{\rm in}\right>$ being the momentum transfer rate coefficient for ion-neutral collisions, $\mathbf{v}_{\rm d}$ is the drift velocity defined as $\mathbf{v}_{\rm d}\equiv\mathbf{u}_{\rm i}-\mathbf{u}_{\rm n}$, $\rho$, $m$, and $\mathbf{u}$ are the density, mass and velocity with subscript ${\rm n}$ and ${\rm i}$ denoting the neutrals and ions, respectively. We utilize the practical drag coefficient $\gamma=3.5\times 10^{13}\,\rm{cm^3\,g^{-1}\second}$ and $\rho_{\rm i}=C\rho_{\rm n}^{1/2}$ with $C=9\times 10^{-16}{\,\rm cm^{-3/2}\, g^{1/2}}$ (three times the value from \citet{Shu1992} but well within the expected range of variation of $C$). Since the drag force on ions $-\mathbf{f}_{\rm d}$ equals negative Lorentz force exerted to ions under the assumption of small fractional ionization, the drift velocity can be rearranged as $\mathbf{v}_{\rm d}=\frac{1}{4\pi\gamma\rho_{\rm n}\rho_{\rm i}}(\nabla\times\mathbf{B})\times\mathbf{B}$. With the strong coupling limit, single fluid is an adequate approximation. Thus the equations remain closed.

\begin{figure*}[htb]
	\centering
    \plottwo{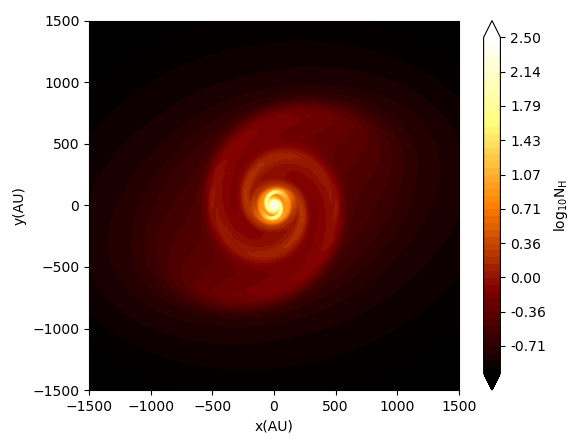}{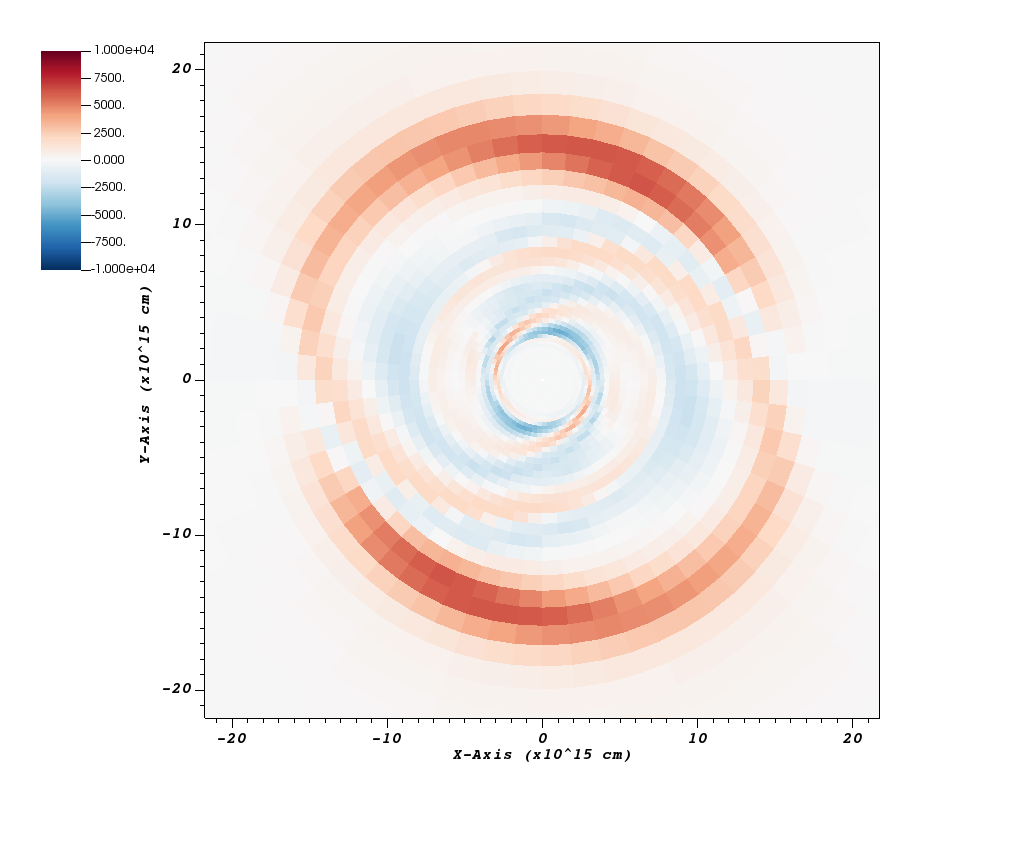}
	\caption{Left: Face-on view of ${\rm log_{10}}$ of column density for model Galpha at $135.0\kyr$. Right: radial drift velocity in the unit of ${\rm cm}\second^{-1}$ on the equatorial plane for model Galpha at $135.0\kyr$. The viewing phase angle of the right panel is rotated by $126\degree$.} \label{fig:Galpha_dv1}
\end{figure*}

We visualize the column density and drift velocity $\mathbf{v}_{\rm d}$ in Figure \ref{fig:Galpha_dv1} to investigate the role that ambipolar diffusion plays in the emergence of the spiral structure. The significant change in column density of model Galpha compared to model G is that the small-scale spiral structure stays two-armed and symmetric in the inner disk. The large-scale magnetic spirals are unaffected in their morphology. 

The drift velocity tends to vanish within $r\sim200\au$ mainly due to high density. In the outer region the drift velocity coincides with the spiral structures. The evolution of drift velocity corresponds to the inside-out collapse and the influential region expands. The interface between positive and negative values of $\mathbf{v}_{\rm d}$ goes along where the magnetic field reverses. Specifically, the positive radial drift velocity (reddish color on the right panel of Figure \ref{fig:Galpha_dv1}) represents where the drag force accelerates the neutrals outwards.

\begin{figure*}[htb]
	\centering
	\includegraphics[width=\textwidth]{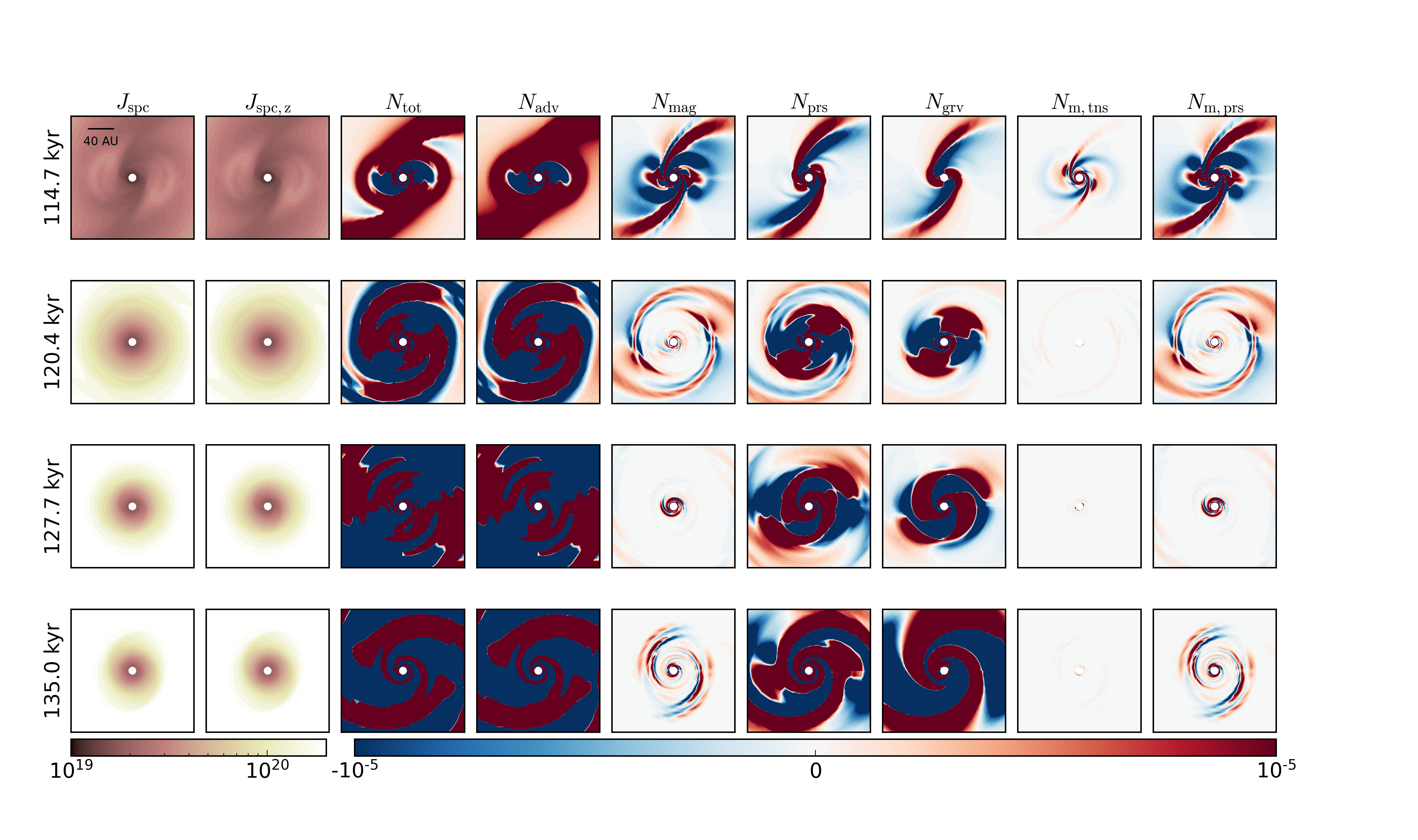}
	\includegraphics[width=\textwidth]{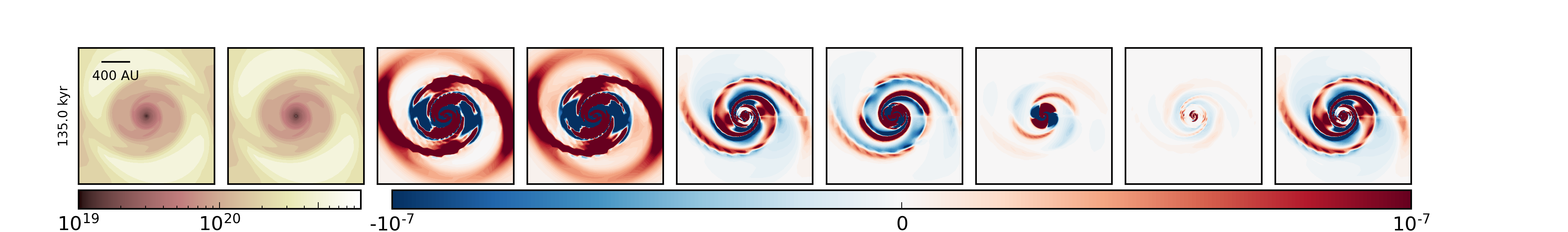}
	\caption{Same as Figure \ref{fig:g5} but for model Galpha}\label{fig:ga5}
\end{figure*}

\begin{figure*}[htb]
	\centering
	\includegraphics[width=\textwidth]{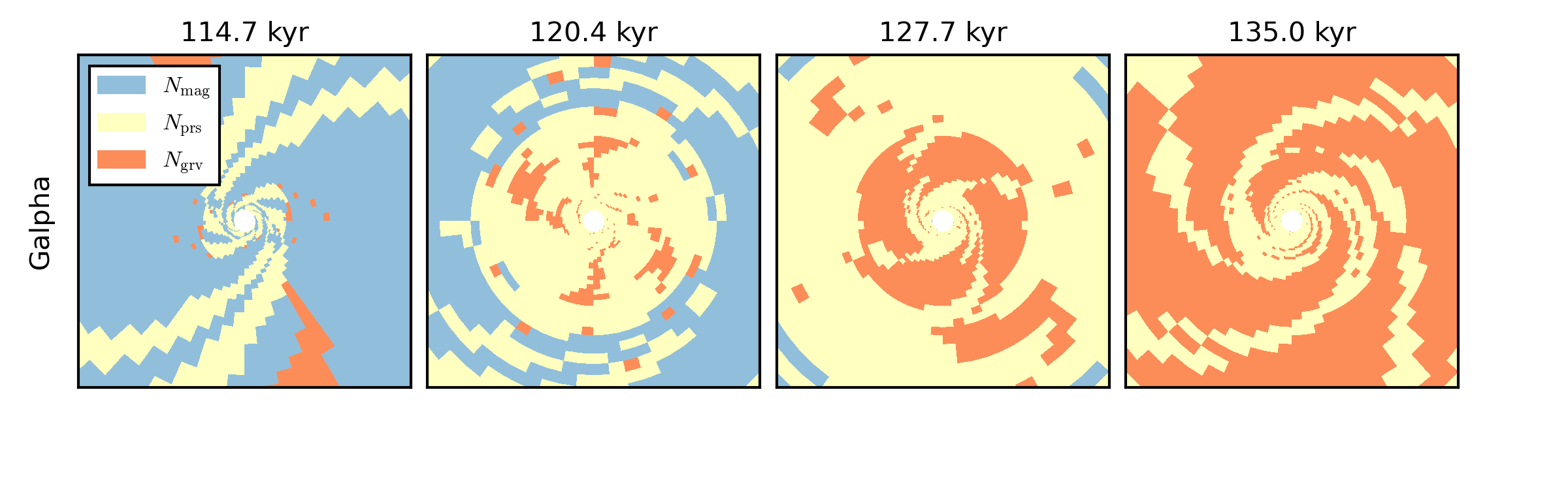}
	\caption{Same as Figure \ref{fig:domTerm} but for model Galpha}\label{fig:domTermga5}
\end{figure*}

The angular momentum and torques of model Galpha are shown in Figure \ref{fig:ga5}. Compared to model G, the major change observed in model Galpha would be the absence of one-armed spiral in torques at later stage but two-armed spirals appear instead, similar to the structure in column density. What remains the same in principle is that at later stage in the most of the inner disk the gravitational torque is prominent and the pressure gradient torque dominates along the spiral edge (see Figure \ref{fig:domTermga5}).

Influences of other aspects on the magnetic spirals, magnetic flux for instance, are beyond the scope of this paper and might be investigated in future work.

\section{Discussion}\label{sec:disc}

\subsection{Gravitational instability affects magnetic field}

Comparing model G with model A, misaligned magnetic fields makes it possible for both gravitational torque and pressure gradient torque to play a role in redistributing angular momentum. Stronger field (model H) or less misalignment (model D), however, leads to less disk mass which is below the threshold of disk mass for gravitational instability to occur. Therefore, both the misalignment of magnetic field and the strength of the field are relevant factors for allowing gravitational instability to grow.

In model G, the wobbling disk leads to both a high mass accretion rate and redistribution of mass. 
Once the rate of mass accumulation is higher than that of magnetic flux, the disk mass is large enough to trigger gravitational instability 
delaying magnetic field gathering.
Due to the motions driven by the gravitational instability, the gas in the disk becomes more evenly spread alongside with the azimuthal magnetic field in the disk, effectively slowing down field gathering around the central sink.

On the other hand, in model H we do not see presence of gravitational instability. There flux is able to sufficiently accumulate to create what we suspect to be a magnetic tower jet, removing angular momentum in that way. This is why we see momentary burst of outflow in model H but not in model G. 

\subsection{Properties of magnetic spirals}\label{sec:spiraldisc}
In misalignment models, a large spiral structure forms at the beginning of accretion due to initial rotation and reversed magnetic field lines. At late stage the central region is disrupted by accumulation of mass and magnetic flux while the outer spirals sustain their structure. 
The formation of RSD essentially highlights the magnetic spirals as a feature of infalling envelope.

Along the outer magnetic spirals total torque keeps positive so as to maintain angular momentum of inflows. 
Since the evolution in the central region is complex, the magnetic spirals most effectively affect the accretion of mass downstream along the spiral arms.
Contrast to the small-scale spirals, the large-scale spirals do not produce shocks.
They are merely features of magnetic inflow, reasonably coherent in their large length scale and independent of the disk down stream.

\subsection{Implications for observations: streamers}

In principle the large-scale magnetic spirals, described in Section \ref{sec:spiraldisc} above,  are very visible features in ideal condition. In practice, however, they are likely easily missed. As sparser features, their emission can be relatively weak and therefore remain unresolved. There are, however, some promising observations. 

Magnetic torques can be relevant with the streamer-like objects such as observed by \citet[][Per-emb-2]{Pineda2020} and \citet[][MGM2012 512]{Grant2021}. Also HH211 observations of \citet{Lee2019} indicate a magnetic streamer arm. If the magnetic field is strong enough to be dynamically significant, it would make the best sense that such inflows are aligned with the magnetic field. As such magnetic torques as explored in this study, could become a significant factor of their behaviour during their collapse downward.

To explain this further, the idea of magnetic spiral inflow is not limited to cases of spirals directly and primarily caused by the magnetic field. The idea also includes cases in which the magnetic field efficiently couples with the inflowing gas. Therefore, even if the spiral-like inflow may be caused by some other phenomenon, dynamically important magnetic field is still crucial.

So there are two primary options, either magnetic field is dynamically significant or it is not. If it is not dynamically significant (unlikely), then we do not need to care about any of this. However, as it likely is, that magnetic field is dynamically significant, we have to look into the effects to reach complete understanding.

If a streamer is to propagate in perpendicular direction with respect to the magnetic field, it would not be able to maintains its elongated structure. In the case of magnetically aligned streamer, however, it would naturally follow the inspiralling form. Therefore, it would be reasonable to state the hypothesis that the streamers of \citet{Pineda2020}, \citet{Grant2021} and \citet{Lee2019} are aligned with the field and that the mean magnetic field of the object would be generally aligned perpendicularly to the rotation axis of the system.

\subsubsection{IRAS 18089--1732}

In the domain of high mass star formation, there is a case of the object IRAS 18089--1732 \citep{Sanhueza2021} which demonstrates, very visibly, magnetic field alignment with the inflowing gas. In their ALMA observations, \citet{Sanhueza2021} depict a site of massive star formation where rotational flows are collapsing towards the centre in whirlpool-like manner. In their estimates, gravity dominates the process with rotational and magnetic energies being significantly weaker. \citet{Sanhueza2021} estimate based on magnetic model and observations that the mass-to-flux ratio to be $\lambda \sim 3.61$ and $\lambda \sim 3.2$ respectively. These would be values between our models I and H, though those are not directly comparable, as our models do not examine massive star formation, and the basis for computing values of $\lambda$ is different.  

\citet{Sanhueza2021} show that IRAS 18089--1732 consist on one spiral streamer and two inflow filaments. Based on polarization estimates, the magnetic field is following the spiral geometry of the streamers/filaments and it is substantially toroidal. Based on the seeming alignment of the field, streamers of IRAS 18089--1732 can be a case of magnetic spirals. 

The magnetic field model of \citet{Sanhueza2021} assumes an hourglass poloidal field with an added toroidal component; but our magnetic spiral model is dominated by toroidal features. However, choosing the basic large-scale field model is a choice which will bias the estimate with its assumptions. The depth of the system is not obvious, and therefore either model assumption can be in principle attempted. 

The basis for magnetic spirals has two main arguments. First, the field geometry is appropriate with density and magnetic structures similarly aligned. Second, despite gravity dominating the system globally, magnetic forces can still play a role locally, e.g. by maintaining relative coherence of the inflow spirals. This is possible, because by the estimates of \citet{Sanhueza2021} magnetic, turbulent and rotational energies are roughly equal. Therefore, within the inflow frame, magnetic field could clearly affect secondary types of motions --- as it happens with magnetic inflow spirals in our collapse models.

\subsection{Implications for observations: rings and spirals}
\begin{figure*}[t]
	\centering
	\includegraphics[width=\textwidth]{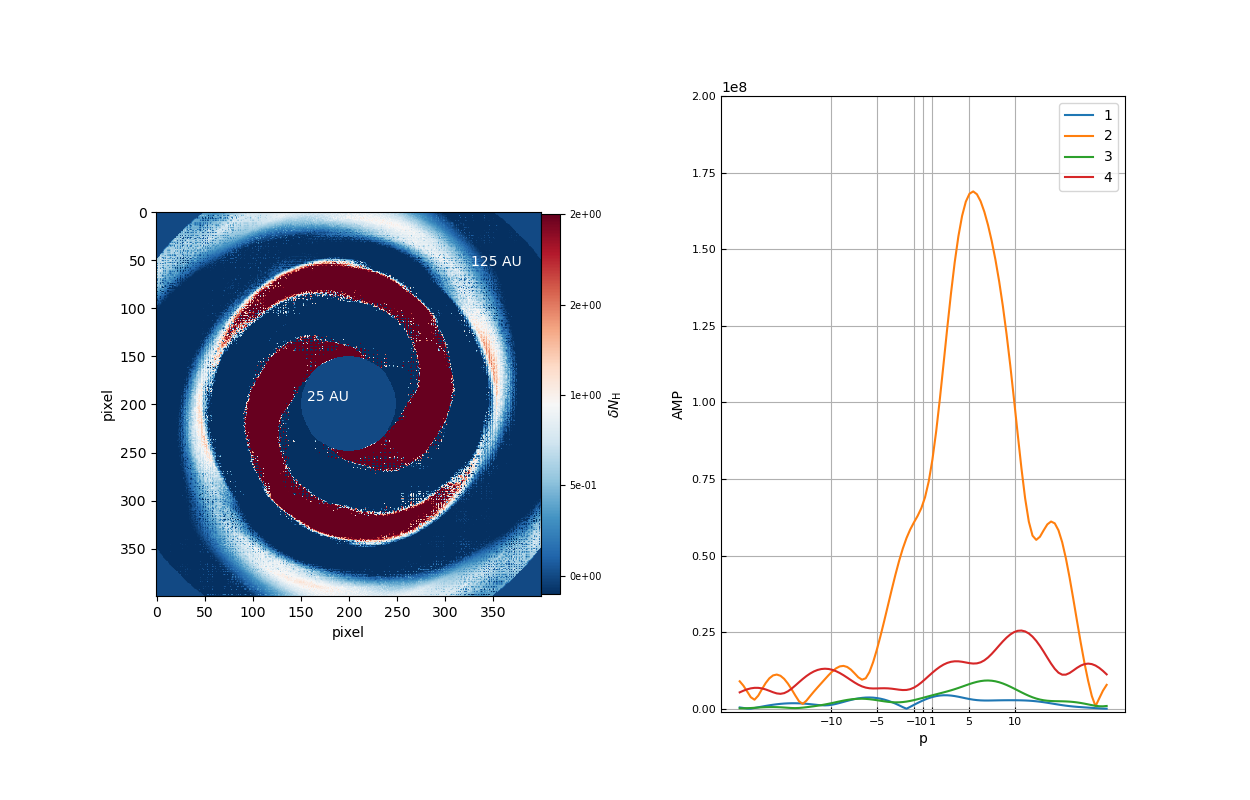}
	\caption{Left panel shows the residual column density $\delta N_{\rm H}$ by subtracting the annular mean from column density of model Galpha at $128.0\kyr$. The cut-off inner and outer radius in pure blue are $25\au$ and $125\au$, respectively. Right panel shows the amplitude of residual column density by 2DDFT. Curves in different colors represent different modes with value of $m$.}\label{fig:Galpha_spiral}
\end{figure*}

A couple of spiral structures within the protostellar disk of HH 111 VLA 1 with ALMA observations of thermal emission have been reported by \citet{Lee2020}. They subtracted the continuum map of HH 111 VLA 1 by its annular mean, and fitted the residual map by logarithmic and Archimedean form, respectively. Their corresponding pitch angles are $\sim16\degree$ for one arm and $\sim13\degree$ for the other. In our simulations, the most promising model to demonstrate the spiral structures in HH 111 VLA 1 is model Galpha, since it has proper length scale (tens of $\au$) and number of arms ($m=2$). To compare with the observed value, we estimate the pitch angle by two-dimensional discrete Fourier transform \citep[2DDFT, e.g.][for details see Appendix \ref{sec:Appendx_2ddft}]{Kalnajs1975,Iye1982,Krakow1982,Yu2018,Yu2019} method more accurately rather than by hand. We processed the map of column density for model Galpha subtracted by its annular mean (Figure \ref{fig:Galpha_spiral}). The pitch angle remains reasonably close from frame to frame. As a result, the dominant mode $(p,m)\approx(6,2)$ corresponds to a pitch angle $\alpha={\rm arctan(\frac{2}{6})}\approx18\degree$, which is consistent with the observed value for HH 111 VLA 1, as long as it can be assumed that the spiral will persist for a longer duration that the numerical collapse simulation has shown.

In Model H, the magnetic rings appear within a warped disk, as two rings which are tilted from each other with respect to their inclinations. This gives us clue for where to find a system with such rings. We found IRAS 04368+2557 could be an interesting one. \citet{Sakai2019} reported evidence of a warped structure in their disk candidate, analogous to two differently aligned rings. It is indicative that disks can be warped and not be constricted into a flat plane. The follow-up observations of the nearly edge-on disk L1527 IRS around the protostar, IRAS 04368+2557, suggest the disk could potentially embed rings in the disk-forming stage \citep{Nakatani2020}. The observed three clumps in the $7\,{\rm mm}$ radio continuum observations are closely located and symmetric, and are resolved in the inner part of the disk ($r<50\au$). \citet{Nakatani2020} speculated it is projected dust ring or spiral arms. Magnetic field provides one natural mechanism how disk could be warped. With continuing modelling work, we could better characterize the magnetic rings, and provide more detailed estimates for observations.

\section{Summary}\label{sec:summary}
In this work, we performed studies of spiral structure in simulations focused on the details of physics. Magnetic spirals in different scales are identified by their morphologies and torques in misaligned models. In some cases with relatively weak magnetic fields, small-scale spirals for which gravitational instability might play a role can be noticed as well. The main results are summarized as follows:

1. Magnetic spirals are triggered by initial rotation where the magnetic inflow aligns itself with the wrapped around magnetic field geometry as it is the most optimal for the flow. 

2. In the misaligned case with a relatively high mass-to-flux ratio (model G) the magnetic torque is the main reason for angular momentum transfer at an early stage. In the central region of the disk, gravitational torque softens the effects of magnetic torques at a late stage. With increased magnetic field (model H), the gravitational torque becomes insignificant.

3. Rings and gaps can form in inner disks when a relatively strong magnetic field wraps itself in the misaligned model with an intermediate mass-to-flux ratio (model H). The local gains and losses of angular momentum are local torques which have ripples in the radial direction; these ripples in the torques may be the cause forming such rings and gaps.

4. Ambipolar diffusion as explored in this work has shown slight influence on the small scale spirals but not the main morphologies of magnetic spirals.

\acknowledgments

This work has made use of tools developed by the CHARMS group in ASIAA, and the High Performance Computing Resource in the Core Facility for Advanced Research Computing at Shanghai Astronomical Observatory and TIARA cluster in ASIAA.  W.W. and F.Y. are supported in part by the Natural Science Foundation of China (grant 11633006) and the Key Research Program of Frontier Sciences of CAS (No. QYZDJSSW-SYS008). M.V., H.S., and R.K. acknowledge funding support for Theory within ASIAA from Academia Sinica. H.S. acknowledges grant support from Ministry of Science and Technology (MoST) in Taiwan through 108-2112-M-001-009- and 109-2112-M-001-028-. ZYL is supported in part by NASA 80NSSC18K1095 and NSF AST1815784.
M.V. and H. S. thank the hospitality of Shanghai Astronomical Observatory, Chinese Academy of Sciences, during their visits. 
This research has made use of NASA's Astrophysics Data System Bibliographic Services.
The authors also would like to thank the anonymous referee for the constructive and insightful comments on the manuscript.

\software{Astropy \citep{astropy2013,astropy2018}, Jupyter notebook \citep{Kluyver:2016aa}, Matplotlib \citep{Hunter:2007}, Numpy \citep{harris2020array}, SciPy \citep{2020SciPy-NMeth}, VisIt \citep[Visualization Tool][]{HPV:VisIt}.}

\bibliography{ms}{}

\begin{thebibliography}{}
\expandafter\ifx\csname natexlab\endcsname\relax\def\natexlab#1{#1}\fi
\providecommand{\url}[1]{\href{#1}{#1}}
\providecommand{\dodoi}[1]{doi:~\href{http://doi.org/#1}{\nolinkurl{#1}}}
\providecommand{\doeprint}[1]{\href{http://ascl.net/#1}{\nolinkurl{http://ascl.net/#1}}}
\providecommand{\doarXiv}[1]{\href{https://arxiv.org/abs/#1}{\nolinkurl{https://arxiv.org/abs/#1}}}

\bibitem[{{Allen} {et~al.}(2003){Allen}, {Li}, \& {Shu}}]{ALS2003}
{Allen}, A., {Li}, Z.-Y., \& {Shu}, F.~H. 2003, \apj, 599, 363,
  \dodoi{10.1086/379243}

\bibitem[{{ALMA Partnership} {et~al.}(2015){ALMA Partnership}, {Brogan},
  {P{\'e}rez}, {Hunter}, {Dent}, {Hales}, {Hills}, {Corder}, {Fomalont},
  {Vlahakis}, {Asaki}, {Barkats}, {Hirota}, {Hodge}, {Impellizzeri}, {Kneissl},
  {Liuzzo}, {Lucas}, {Marcelino}, {Matsushita}, {Nakanishi}, {Phillips},
  {Richards}, {Toledo}, {Aladro}, {Broguiere}, {Cortes}, {Cortes}, {Espada},
  {Galarza}, {Garcia-Appadoo}, {Guzman-Ramirez}, {Humphreys}, {Jung}, {Kameno},
  {Laing}, {Leon}, {Marconi}, {Mignano}, {Nikolic}, {Nyman}, {Radiszcz},
  {Remijan}, {Rod{\'o}n}, {Sawada}, {Takahashi}, {Tilanus}, {Vila Vilaro},
  {Watson}, {Wiklind}, {Akiyama}, {Chapillon}, {de Gregorio-Monsalvo}, {Di
  Francesco}, {Gueth}, {Kawamura}, {Lee}, {Nguyen Luong}, {Mangum}, {Pietu},
  {Sanhueza}, {Saigo}, {Takakuwa}, {Ubach}, {van Kempen}, {Wootten},
  {Castro-Carrizo}, {Francke}, {Gallardo}, {Garcia}, {Gonzalez}, {Hill},
  {Kaminski}, {Kurono}, {Liu}, {Lopez}, {Morales}, {Plarre}, {Schieven},
  {Testi}, {Videla}, {Villard}, {Andreani}, {Hibbard}, \&
  {Tatematsu}}]{ALMA2015}
{ALMA Partnership}, {Brogan}, C.~L., {P{\'e}rez}, L.~M., {et~al.} 2015, \apjl,
  808, L3, \dodoi{10.1088/2041-8205/808/1/L3}

\bibitem[{{Andrews} {et~al.}(2018){Andrews}, {Huang}, {P{\'e}rez}, {Isella},
  {Dullemond}, {Kurtovic}, {Guzm{\'a}n}, {Carpenter}, {Wilner}, {Zhang}, {Zhu},
  {Birnstiel}, {Bai}, {Benisty}, {Hughes}, {{\"O}berg}, \&
  {Ricci}}]{DHARPI2018}
{Andrews}, S.~M., {Huang}, J., {P{\'e}rez}, L.~M., {et~al.} 2018, \apjl, 869,
  L41, \dodoi{10.3847/2041-8213/aaf741}

\bibitem[{{Ansdell} {et~al.}(2017){Ansdell}, {Williams}, {Manara}, {Miotello},
  {Facchini}, {van der Marel}, {Testi}, \& {van Dishoeck}}]{Ansdell2017}
{Ansdell}, M., {Williams}, J.~P., {Manara}, C.~F., {et~al.} 2017, \aj, 153,
  240, \dodoi{10.3847/1538-3881/aa69c0}

\bibitem[{{Ansdell} {et~al.}(2016){Ansdell}, {Williams}, {van der Marel},
  {Carpenter}, {Guidi}, {Hogerheijde}, {Mathews}, {Manara}, {Miotello},
  {Natta}, {Oliveira}, {Tazzari}, {Testi}, {van Dishoeck}, \& {van
  Terwisga}}]{Ansdell2016}
{Ansdell}, M., {Williams}, J.~P., {van der Marel}, N., {et~al.} 2016, \apj,
  828, 46, \dodoi{10.3847/0004-637X/828/1/46}

\bibitem[{{Ansdell} {et~al.}(2018){Ansdell}, {Williams}, {Trapman}, {van
  Terwisga}, {Facchini}, {Manara}, {van der Marel}, {Miotello}, {Tazzari},
  {Hogerheijde}, {Guidi}, {Testi}, \& {van Dishoeck}}]{Ansdell2018}
{Ansdell}, M., {Williams}, J.~P., {Trapman}, L., {et~al.} 2018, \apj, 859, 21,
  \dodoi{10.3847/1538-4357/aab890}

\bibitem[{{Astropy Collaboration} {et~al.}(2013){Astropy Collaboration},
  {Robitaille}, {Tollerud}, {Greenfield}, {Droettboom}, {Bray}, {Aldcroft},
  {Davis}, {Ginsburg}, {Price-Whelan}, {Kerzendorf}, {Conley}, {Crighton},
  {Barbary}, {Muna}, {Ferguson}, {Grollier}, {Parikh}, {Nair}, {Unther},
  {Deil}, {Woillez}, {Conseil}, {Kramer}, {Turner}, {Singer}, {Fox}, {Weaver},
  {Zabalza}, {Edwards}, {Azalee Bostroem}, {Burke}, {Casey}, {Crawford},
  {Dencheva}, {Ely}, {Jenness}, {Labrie}, {Lim}, {Pierfederici}, {Pontzen},
  {Ptak}, {Refsdal}, {Servillat}, \& {Streicher}}]{astropy2013}
{Astropy Collaboration}, {Robitaille}, T.~P., {Tollerud}, E.~J., {et~al.} 2013,
  \aap, 558, A33, \dodoi{10.1051/0004-6361/201322068}

\bibitem[{{Astropy Collaboration} {et~al.}(2018){Astropy Collaboration},
  {Price-Whelan}, {Sip{\H{o}}cz}, {G{\"u}nther}, {Lim}, {Crawford}, {Conseil},
  {Shupe}, {Craig}, {Dencheva}, {Ginsburg}, {VanderPlas}, {Bradley},
  {P{\'e}rez-Su{\'a}rez}, {de Val-Borro}, {Aldcroft}, {Cruz}, {Robitaille},
  {Tollerud}, {Ardelean}, {Babej}, {Bach}, {Bachetti}, {Bakanov}, {Bamford},
  {Barentsen}, {Barmby}, {Baumbach}, {Berry}, {Biscani}, {Boquien}, {Bostroem},
  {Bouma}, {Brammer}, {Bray}, {Breytenbach}, {Buddelmeijer}, {Burke},
  {Calderone}, {Cano Rodr{\'\i}guez}, {Cara}, {Cardoso}, {Cheedella}, {Copin},
  {Corrales}, {Crichton}, {D'Avella}, {Deil}, {Depagne}, {Dietrich}, {Donath},
  {Droettboom}, {Earl}, {Erben}, {Fabbro}, {Ferreira}, {Finethy}, {Fox},
  {Garrison}, {Gibbons}, {Goldstein}, {Gommers}, {Greco}, {Greenfield},
  {Groener}, {Grollier}, {Hagen}, {Hirst}, {Homeier}, {Horton}, {Hosseinzadeh},
  {Hu}, {Hunkeler}, {Ivezi{\'c}}, {Jain}, {Jenness}, {Kanarek}, {Kendrew},
  {Kern}, {Kerzendorf}, {Khvalko}, {King}, {Kirkby}, {Kulkarni}, {Kumar},
  {Lee}, {Lenz}, {Littlefair}, {Ma}, {Macleod}, {Mastropietro}, {McCully},
  {Montagnac}, {Morris}, {Mueller}, {Mumford}, {Muna}, {Murphy}, {Nelson},
  {Nguyen}, {Ninan}, {N{\"o}the}, {Ogaz}, {Oh}, {Parejko}, {Parley}, {Pascual},
  {Patil}, {Patil}, {Plunkett}, {Prochaska}, {Rastogi}, {Reddy Janga},
  {Sabater}, {Sakurikar}, {Seifert}, {Sherbert}, {Sherwood-Taylor}, {Shih},
  {Sick}, {Silbiger}, {Singanamalla}, {Singer}, {Sladen}, {Sooley},
  {Sornarajah}, {Streicher}, {Teuben}, {Thomas}, {Tremblay}, {Turner},
  {Terr{\'o}n}, {van Kerkwijk}, {de la Vega}, {Watkins}, {Weaver}, {Whitmore},
  {Woillez}, {Zabalza}, \& {Astropy Contributors}}]{astropy2018}
{Astropy Collaboration}, {Price-Whelan}, A.~M., {Sip{\H{o}}cz}, B.~M., {et~al.}
  2018, \aj, 156, 123, \dodoi{10.3847/1538-3881/aabc4f}

\bibitem[{{Carpenter} {et~al.}(2014){Carpenter}, {Ricci}, \&
  {Isella}}]{Carpenter2014}
{Carpenter}, J.~M., {Ricci}, L., \& {Isella}, A. 2014, \apj, 787, 42,
  \dodoi{10.1088/0004-637X/787/1/42}

\bibitem[{{Cazzoletti} {et~al.}(2019){Cazzoletti}, {Manara}, {Baobab Liu}, {van
  Dishoeck}, {Facchini}, {Alcal{\`a}}, {Ansdell}, {Testi}, {Williams},
  {Carrasco-Gonz{\'a}lez}, {Dong}, {Forbrich}, {Fukagawa}, {Galv{\'a}n-Madrid},
  {Hirano}, {Hogerheijde}, {Hasegawa}, {Muto}, {Pinilla}, {Takami}, {Tamura},
  {Tazzari}, \& {Wisniewski}}]{Cazzoletti2019}
{Cazzoletti}, P., {Manara}, C.~F., {Baobab Liu}, H., {et~al.} 2019, \aap, 626,
  A11, \dodoi{10.1051/0004-6361/201935273}

\bibitem[{Childs {et~al.}(2012)Childs, Brugger, Whitlock, Meredith, Ahern,
  Pugmire, Biagas, Miller, Harrison, Weber, Krishnan, Fogal, Sanderson, Garth,
  Bethel, Camp, R\"{u}bel, Durant, Favre, \& Navr\'{a}til}]{HPV:VisIt}
Childs, H., Brugger, E., Whitlock, B., {et~al.} 2012, in High Performance
  Visualization--Enabling Extreme-Scale Scientific Insight, 357--372

\bibitem[{{Cieza} {et~al.}(2021){Cieza}, {Gonz{\'a}lez-Ruilova}, {Hales},
  {Pinilla}, {Ru{\'\i}z-Rodr{\'\i}guez}, {Zurlo}, {Casassus}, {P{\'e}rez},
  {C{\'a}novas}, {Arce-Tord}, {Flock}, {Kurtovic}, {Marino}, {Nogueira},
  {Perez}, {Price}, {Principe}, \& {Williams}}]{Cieza2021}
{Cieza}, L.~A., {Gonz{\'a}lez-Ruilova}, C., {Hales}, A.~S., {et~al.} 2021,
  \mnras, 501, 2934, \dodoi{10.1093/mnras/staa3787}

\bibitem[{{Das} \& {Basu}(2021)}]{DB2021}
{Das}, I., \& {Basu}, S. 2021, \apj, 910, 163, \dodoi{10.3847/1538-4357/abdb2c}

\bibitem[{{Duffin} \& {Pudritz}(2008)}]{Duffin2008}
{Duffin}, D.~F., \& {Pudritz}, R.~E. 2008, \mnras, 391, 1659,
  \dodoi{10.1111/j.1365-2966.2008.14026.x}

\bibitem[{{Dutta} {et~al.}(2020){Dutta}, {Lee}, {Liu}, {Hirano}, {Liu},
  {Tatematsu}, {Kim}, {Shang}, {Sahu}, {Kim}, {Moraghan}, {Jhan}, {Hsu},
  {Evans}, {Johnstone}, {Ward-Thompson}, {Kuan}, {Lee}, {Lee}, {Traficante},
  {Juvela}, {Vastel}, {Zhang}, {Sanhueza}, {Soam}, {Kwon}, {Bronfman}, {Eden},
  {Goldsmith}, {He}, {Wu}, {Pelkonen}, {Qin}, {Li}, \& {Li}}]{Dutta2020}
{Dutta}, S., {Lee}, C.-F., {Liu}, T., {et~al.} 2020, \apjs, 251, 20,
  \dodoi{10.3847/1538-4365/abba26}

\bibitem[{{Galli} {et~al.}(2006){Galli}, {Lizano}, {Shu}, \&
  {Allen}}]{GLSA2006}
{Galli}, D., {Lizano}, S., {Shu}, F.~H., \& {Allen}, A. 2006, \apj, 647, 374,
  \dodoi{10.1086/505257}

\bibitem[{{Goldreich} \& {Lynden-Bell}(1965)}]{Goldreich1965}
{Goldreich}, P., \& {Lynden-Bell}, D. 1965, \mnras, 130, 97,
  \dodoi{10.1093/mnras/130.2.97}

\bibitem[{{Grant} {et~al.}(2021){Grant}, {Espaillat}, {Wendeborn}, {Tobin},
  {Mac{\'\i}as}, {Rilinger}, {Ribas}, {Megeath}, {Fischer}, {Calvet}, \& {Hee
  Kim}}]{Grant2021}
{Grant}, S.~L., {Espaillat}, C.~C., {Wendeborn}, J., {et~al.} 2021, \apj, 913,
  123, \dodoi{10.3847/1538-4357/abf432}

\bibitem[{Harris {et~al.}(2020)Harris, Millman, van~der Walt, Gommers,
  Virtanen, Cournapeau, Wieser, Taylor, Berg, Smith, Kern, Picus, Hoyer, van
  Kerkwijk, Brett, Haldane, del R{\'{i}}o, Wiebe, Peterson,
  G{\'{e}}rard-Marchant, Sheppard, Reddy, Weckesser, Abbasi, Gohlke, \&
  Oliphant}]{harris2020array}
Harris, C.~R., Millman, K.~J., van~der Walt, S.~J., {et~al.} 2020, Nature, 585,
  357, \dodoi{10.1038/s41586-020-2649-2}

\bibitem[{{Harsono} {et~al.}(2021){Harsono}, {van der Wiel}, {Bjerkeli},
  {Ramsey}, {Calcutt}, {Kristensen}, \& {J{\o}rgensen}}]{Harsono2021}
{Harsono}, D., {van der Wiel}, M.~H.~D., {Bjerkeli}, P., {et~al.} 2021, \aap,
  646, A72, \dodoi{10.1051/0004-6361/202038697}

\bibitem[{{Hennebelle} \& {Ciardi}(2009)}]{Hennebelle2009}
{Hennebelle}, P., \& {Ciardi}, A. 2009, \aap, 506, L29,
  \dodoi{10.1051/0004-6361/200913008}

\bibitem[{{Hennebelle} {et~al.}(2016){Hennebelle}, {Commer{\c{c}}on},
  {Chabrier}, \& {Marchand}}]{Hennebelle2016}
{Hennebelle}, P., {Commer{\c{c}}on}, B., {Chabrier}, G., \& {Marchand}, P.
  2016, \apjl, 830, L8, \dodoi{10.3847/2041-8205/830/1/L8}

\bibitem[{{Hsu} {et~al.}(2020){Hsu}, {Liu}, {Liu}, {Sahu}, {Hirano}, {Lee},
  {Tatematsu}, {Kim}, {Juvela}, {Sanhueza}, {He}, {Johnstone}, {Qin},
  {Bronfman}, {Chen}, {Dutta}, {Eden}, {Jhan}, {Kim}, {Kuan}, {Kwon}, {Lee},
  {Lee}, {Moraghan}, {Rawlings}, {Shang}, {Soam}, {Thompson}, {Traficante},
  {Wu}, {Yang}, \& {Zhang}}]{Hsu2020}
{Hsu}, S.-Y., {Liu}, S.-Y., {Liu}, T., {et~al.} 2020, \apj, 898, 107,
  \dodoi{10.3847/1538-4357/ab9f3a}

\bibitem[{Hunter(2007)}]{Hunter:2007}
Hunter, J.~D. 2007, Computing in Science \& Engineering, 9, 90,
  \dodoi{10.1109/MCSE.2007.55}

\bibitem[{{Iye} {et~al.}(1982){Iye}, {Okamura}, {Hamabe}, \&
  {Watanabe}}]{Iye1982}
{Iye}, M., {Okamura}, S., {Hamabe}, M., \& {Watanabe}, M. 1982, \apj, 256, 103,
  \dodoi{10.1086/159887}

\bibitem[{{Johns-Krull} {et~al.}(2004){Johns-Krull}, {Valenti}, \&
  {Saar}}]{JohnsKrull2004}
{Johns-Krull}, C.~M., {Valenti}, J.~A., \& {Saar}, S.~H. 2004, \apj, 617, 1204,
  \dodoi{10.1086/425652}

\bibitem[{{Joos} {et~al.}(2012){Joos}, {Hennebelle}, \& {Ciardi}}]{JHC2012}
{Joos}, M., {Hennebelle}, P., \& {Ciardi}, A. 2012, \aap, 543, A128,
  \dodoi{10.1051/0004-6361/201118730}

\bibitem[{{Kalnajs}(1975)}]{Kalnajs1975}
{Kalnajs}, A.~J. 1975, in La Dynamique des galaxies spirales, ed.
  L.~{Weliachew}, Vol. 241, 103

\bibitem[{Kluyver {et~al.}(2016)Kluyver, Ragan-Kelley, P{\'e}rez, Granger,
  Bussonnier, Frederic, Kelley, Hamrick, Grout, Corlay, Ivanov, Avila, Abdalla,
  \& Willing}]{Kluyver:2016aa}
Kluyver, T., Ragan-Kelley, B., P{\'e}rez, F., {et~al.} 2016, in Positioning and
  Power in Academic Publishing: Players, Agents and Agendas, ed. F.~Loizides \&
  B.~Schmidt, IOS Press, 87 -- 90.
\newblock \url{https://eprints.soton.ac.uk/403913/}

\bibitem[{{Krakow} {et~al.}(1982){Krakow}, {Huntley}, \& {Seiden}}]{Krakow1982}
{Krakow}, W., {Huntley}, J.~M., \& {Seiden}, P.~E. 1982, \aj, 87, 203,
  \dodoi{10.1086/113097}

\bibitem[{{Krasnopolsky} {et~al.}(2010){Krasnopolsky}, {Li}, \&
  {Shang}}]{KLS2010}
{Krasnopolsky}, R., {Li}, Z.-Y., \& {Shang}, H. 2010, \apj, 716, 1541,
  \dodoi{10.1088/0004-637X/716/2/1541}

\bibitem[{{Kratter} \& {Lodato}(2016)}]{Kratter2016}
{Kratter}, K., \& {Lodato}, G. 2016, \araa, 54, 271,
  \dodoi{10.1146/annurev-astro-081915-023307}

\bibitem[{{Lee} {et~al.}(2020){Lee}, {Li}, \& {Turner}}]{Lee2020}
{Lee}, C.-F., {Li}, Z.-Y., \& {Turner}, N.~J. 2020, Nature Astronomy, 4, 142,
  \dodoi{10.1038/s41550-019-0905-x}

\bibitem[{{Lee} {et~al.}(2019){Lee}, {Kwon}, {Jhan}, {Hirano}, {Hwang}, {Lai},
  {Ching}, {Rao}, \& {Ho}}]{Lee2019}
{Lee}, C.-F., {Kwon}, W., {Jhan}, K.-S., {et~al.} 2019, \apj, 879, 101,
  \dodoi{10.3847/1538-4357/ab2458}

\bibitem[{{Li} {et~al.}(2013){Li}, {Krasnopolsky}, \& {Shang}}]{LKS2013}
{Li}, Z.-Y., {Krasnopolsky}, R., \& {Shang}, H. 2013, \apj, 774, 82,
  \dodoi{10.1088/0004-637X/774/1/82}

\bibitem[{{Lin} \& {Shu}(1964)}]{LinShu1964}
{Lin}, C.~C., \& {Shu}, F.~H. 1964, \apj, 140, 646, \dodoi{10.1086/147955}

\bibitem[{{Lizano} {et~al.}(2010){Lizano}, {Galli}, {Cai}, \&
  {Adams}}]{LGCA2010}
{Lizano}, S., {Galli}, D., {Cai}, M.~J., \& {Adams}, F.~C. 2010, \apj, 724,
  1561, \dodoi{10.1088/0004-637X/724/2/1561}

\bibitem[{{Long} {et~al.}(2017){Long}, {Herczeg}, {Pascucci}, {Drabek-Maunder},
  {Mohanty}, {Testi}, {Apai}, {Hendler}, {Henning}, {Manara}, \&
  {Mulders}}]{Long2017}
{Long}, F., {Herczeg}, G.~J., {Pascucci}, I., {et~al.} 2017, \apj, 844, 99,
  \dodoi{10.3847/1538-4357/aa78fc}

\bibitem[{{Long} {et~al.}(2018{\natexlab{a}}){Long}, {Herczeg}, {Pascucci},
  {Apai}, {Henning}, {Manara}, {Mulders}, {Sz{\H{u}}cs}, \&
  {Hendler}}]{Long2018a}
---. 2018{\natexlab{a}}, \apj, 863, 61, \dodoi{10.3847/1538-4357/aacce9}

\bibitem[{{Long} {et~al.}(2018{\natexlab{b}}){Long}, {Pinilla}, {Herczeg},
  {Harsono}, {Dipierro}, {Pascucci}, {Hendler}, {Tazzari}, {Ragusa}, {Salyk},
  {Edwards}, {Lodato}, {van de Plas}, {Johnstone}, {Liu}, {Boehler}, {Cabrit},
  {Manara}, {Menard}, {Mulders}, {Nisini}, {Fischer}, {Rigliaco}, {Banzatti},
  {Avenhaus}, \& {Gully-Santiago}}]{Long2018}
{Long}, F., {Pinilla}, P., {Herczeg}, G.~J., {et~al.} 2018{\natexlab{b}}, \apj,
  869, 17, \dodoi{10.3847/1538-4357/aae8e1}

\bibitem[{{Long} {et~al.}(2019){Long}, {Herczeg}, {Harsono}, {Pinilla},
  {Tazzari}, {Manara}, {Pascucci}, {Cabrit}, {Nisini}, {Johnstone}, {Edwards},
  {Salyk}, {Menard}, {Lodato}, {Boehler}, {Mace}, {Liu}, {Mulders}, {Hendler},
  {Ragusa}, {Fischer}, {Banzatti}, {Rigliaco}, {van de Plas}, {Dipierro},
  {Gully-Santiago}, \& {Lopez-Valdivia}}]{Long2019}
{Long}, F., {Herczeg}, G.~J., {Harsono}, D., {et~al.} 2019, \apj, 882, 49,
  \dodoi{10.3847/1538-4357/ab2d2d}

\bibitem[{{Marchand} {et~al.}(2016){Marchand}, {Masson}, {Chabrier},
  {Hennebelle}, {Commer{\c{c}}on}, \& {Vaytet}}]{Marchand2016}
{Marchand}, P., {Masson}, J., {Chabrier}, G., {et~al.} 2016, \aap, 592, A18,
  \dodoi{10.1051/0004-6361/201526780}

\bibitem[{{Masson} {et~al.}(2016){Masson}, {Chabrier}, {Hennebelle}, {Vaytet},
  \& {Commer{\c{c}}on}}]{Masson2016}
{Masson}, J., {Chabrier}, G., {Hennebelle}, P., {Vaytet}, N., \&
  {Commer{\c{c}}on}, B. 2016, \aap, 587, A32,
  \dodoi{10.1051/0004-6361/201526371}

\bibitem[{{Mellon} \& {Li}(2009)}]{Mellon2009}
{Mellon}, R.~R., \& {Li}, Z.-Y. 2009, \apj, 698, 922,
  \dodoi{10.1088/0004-637X/698/1/922}

\bibitem[{{Mestel} \& {Spitzer}(1956)}]{mestel1956}
{Mestel}, L., \& {Spitzer}, L., J. 1956, \mnras, 116, 503,
  \dodoi{10.1093/mnras/116.5.503}

\bibitem[{{Nakatani} {et~al.}(2020){Nakatani}, {Liu}, {Ohashi}, {Zhang},
  {Hanawa}, {Chandler}, {Oya}, \& {Sakai}}]{Nakatani2020}
{Nakatani}, R., {Liu}, H.~B., {Ohashi}, S., {et~al.} 2020, \apjl, 895, L2,
  \dodoi{10.3847/2041-8213/ab8eaa}

\bibitem[{{Pineda} {et~al.}(2020){Pineda}, {Segura-Cox}, {Caselli},
  {Cunningham}, {Zhao}, {Schmiedeke}, {Maureira}, \& {Neri}}]{Pineda2020}
{Pineda}, J.~E., {Segura-Cox}, D., {Caselli}, P., {et~al.} 2020, Nature
  Astronomy, 4, 1158, \dodoi{10.1038/s41550-020-1150-z}

\bibitem[{{Podio} {et~al.}(2020){Podio}, {Garufi}, {Codella}, {Fedele},
  {Bianchi}, {Bacciotti}, {Ceccarelli}, {Favre}, {Mercimek}, {Rygl}, \&
  {Testi}}]{Podio2020}
{Podio}, L., {Garufi}, A., {Codella}, C., {et~al.} 2020, \aap, 642, L7,
  \dodoi{10.1051/0004-6361/202038952}

\bibitem[{{Ru{\'\i}z-Rodr{\'\i}guez} {et~al.}(2018){Ru{\'\i}z-Rodr{\'\i}guez},
  {Cieza}, {Williams}, {Andrews}, {Principe}, {Caceres}, {Canovas}, {Casassus},
  {Schreiber}, \& {Kastner}}]{ruiz2018}
{Ru{\'\i}z-Rodr{\'\i}guez}, D., {Cieza}, L.~A., {Williams}, J.~P., {et~al.}
  2018, \mnras, 478, 3674, \dodoi{10.1093/mnras/sty1351}

\bibitem[{{Sadavoy} {et~al.}(2019){Sadavoy}, {Stephens}, {Myers}, {Looney},
  {Tobin}, {Kwon}, {Commer{\c{c}}on}, {Segura-Cox}, {Henning}, \&
  {Hennebelle}}]{Sadavoy2019}
{Sadavoy}, S.~I., {Stephens}, I.~W., {Myers}, P.~C., {et~al.} 2019, \apjs, 245,
  2, \dodoi{10.3847/1538-4365/ab4257}

\bibitem[{{Sahu} {et~al.}(2021){Sahu}, {Liu}, {Liu}, {Evans}, {Hirano},
  {Tatematsu}, {Lee}, {Kim}, {Dutta}, {Alina}, {Bronfman}, {Cunningham},
  {Eden}, {Garay}, {Goldsmith}, {He}, {Hsu}, {Jhan}, {Johnstone}, {Juvela},
  {Kim}, {Kuan}, {Kwon}, {Lee}, {Lee}, {Li}, {Li}, {Li}, {Luo}, {Montillaud},
  {Moraghan}, {Pelkonen}, {Qin}, {Ristorcelli}, {Sanhueza}, {Shang}, {Shen},
  {Soam}, {Wu}, {Zhang}, \& {Zhou}}]{Sahu2021}
{Sahu}, D., {Liu}, S.-Y., {Liu}, T., {et~al.} 2021, \apjl, 907, L15,
  \dodoi{10.3847/2041-8213/abd3aa}

\bibitem[{{Sakai} {et~al.}(2019){Sakai}, {Hanawa}, {Zhang}, {Higuchi},
  {Ohashi}, {Oya}, \& {Yamamoto}}]{Sakai2019}
{Sakai}, N., {Hanawa}, T., {Zhang}, Y., {et~al.} 2019, \nat, 565, 206,
  \dodoi{10.1038/s41586-018-0819-2}

\bibitem[{{Sanhueza} {et~al.}(2021){Sanhueza}, {Girart}, {Padovani}, {Galli},
  {Hull}, {Zhang}, {Cortes}, {Stephens}, {Fern{\'a}ndez-L{\'o}pez}, {Jackson},
  {Frau}, {Kock}, {Wu}, {Zapata}, {Olguin}, {Lu}, {Silva}, {Tang}, {Sakai},
  {Guzm{\'a}n}, {Tatematsu}, {Nakamura}, \& {Chen}}]{Sanhueza2021}
{Sanhueza}, P., {Girart}, J.~M., {Padovani}, M., {et~al.} 2021, \apjl, 915,
  L10, \dodoi{10.3847/2041-8213/ac081c}

\bibitem[{{Shang} {et~al.}(2020){Shang}, {Krasnopolsky}, {Liu}, \&
  {Wang}}]{Shang2020}
{Shang}, H., {Krasnopolsky}, R., {Liu}, C.-F., \& {Wang}, L.-Y. 2020, \apj,
  905, 116, \dodoi{10.3847/1538-4357/abbdb0}

\bibitem[{{Sheehan} {et~al.}(2020){Sheehan}, {Tobin}, {Federman}, {Megeath}, \&
  {Looney}}]{Sheehan2020}
{Sheehan}, P.~D., {Tobin}, J.~J., {Federman}, S., {Megeath}, S.~T., \&
  {Looney}, L.~W. 2020, \apj, 902, 141, \dodoi{10.3847/1538-4357/abbad5}

\bibitem[{{Shu}(1992)}]{Shu1992}
{Shu}, F.~H. 1992, {Physics of Astrophysics, Vol. II}

\bibitem[{{Shu} \& {Li}(1997)}]{SL1997}
{Shu}, F.~H., \& {Li}, Z.-Y. 1997, \apj, 475, 251, \dodoi{10.1086/303521}

\bibitem[{{Shu} {et~al.}(1995){Shu}, {Najita}, {Ostriker}, \&
  {Shang}}]{SNOS1995}
{Shu}, F.~H., {Najita}, J., {Ostriker}, E.~C., \& {Shang}, H. 1995, \apjl, 455,
  L155, \dodoi{10.1086/309838}

\bibitem[{{Tazzari} {et~al.}(2021){Tazzari}, {Testi}, {Natta}, {Williams},
  {Ansdell}, {Carpenter}, {Facchini}, {Guidi}, {Hogherheijde}, {Manara},
  {Miotello}, \& {van der Marel}}]{Tazzari2020}
{Tazzari}, M., {Testi}, L., {Natta}, A., {et~al.} 2021, \mnras, 506, 5117,
  \dodoi{10.1093/mnras/stab1912}

\bibitem[{{Tobin} {et~al.}(2019){Tobin}, {Megeath}, {van't Hoff},
  {D{\'\i}az-Rodr{\'\i}guez}, {Reynolds}, {Osorio}, {Anglada}, {Furlan},
  {Karnath}, {Offner}, {Sheehan}, {Sadavoy}, {Stutz}, {Fischer}, {Kama},
  {Persson}, {Di Francesco}, {Looney}, {Watson}, {Li}, {Stephens}, {Chandler},
  {Cox}, {Dunham}, {Kratter}, {Kounkel}, {Mazur}, {Murillo}, {Patel}, {Perez},
  {Segura-Cox}, {Sharma}, {Tychoniec}, \& {Wyrowski}}]{Tobin2019}
{Tobin}, J.~J., {Megeath}, S.~T., {van't Hoff}, M., {et~al.} 2019, \apj, 886,
  6, \dodoi{10.3847/1538-4357/ab498f}

\bibitem[{{Tobin} {et~al.}(2020){Tobin}, {Sheehan}, {Megeath},
  {D{\'\i}az-Rodr{\'\i}guez}, {Offner}, {Murillo}, {van 't Hoff}, {van
  Dishoeck}, {Osorio}, {Anglada}, {Furlan}, {Stutz}, {Reynolds}, {Karnath},
  {Fischer}, {Persson}, {Looney}, {Li}, {Stephens}, {Chandler}, {Cox},
  {Dunham}, {Tychoniec}, {Kama}, {Kratter}, {Kounkel}, {Mazur}, {Maud},
  {Patel}, {Perez}, {Sadavoy}, {Segura-Cox}, {Sharma}, {Stephenson}, {Watson},
  \& {Wyrowski}}]{Tobin2020}
{Tobin}, J.~J., {Sheehan}, P.~D., {Megeath}, S.~T., {et~al.} 2020, \apj, 890,
  130, \dodoi{10.3847/1538-4357/ab6f64}

\bibitem[{{Toomre}(1964)}]{Toomre1964}
{Toomre}, A. 1964, \apj, 139, 1217, \dodoi{10.1086/147861}

\bibitem[{{Tsukamoto} {et~al.}(2015){Tsukamoto}, {Iwasaki}, {Okuzumi},
  {Machida}, \& {Inutsuka}}]{Tsukamoto2015}
{Tsukamoto}, Y., {Iwasaki}, K., {Okuzumi}, S., {Machida}, M.~N., \& {Inutsuka},
  S. 2015, \mnras, 452, 278, \dodoi{10.1093/mnras/stv1290}

\bibitem[{{V{\"a}is{\"a}l{\"a}} {et~al.}(2019){V{\"a}is{\"a}l{\"a}}, {Shang},
  {Krasnopolsky}, {Liu}, {Lam}, \& {Li}}]{MV2019}
{V{\"a}is{\"a}l{\"a}}, M.~S., {Shang}, H., {Krasnopolsky}, R., {et~al.} 2019,
  \apj, 873, 114, \dodoi{10.3847/1538-4357/ab0307}

\bibitem[{{Vaytet} {et~al.}(2018){Vaytet}, {Commer{\c{c}}on}, {Masson},
  {Gonz{\'a}lez}, \& {Chabrier}}]{Vaytet2018}
{Vaytet}, N., {Commer{\c{c}}on}, B., {Masson}, J., {Gonz{\'a}lez}, M., \&
  {Chabrier}, G. 2018, \aap, 615, A5, \dodoi{10.1051/0004-6361/201732075}

\bibitem[{Virtanen {et~al.}(2020)Virtanen, Gommers, Oliphant, Haberland, Reddy,
  Cournapeau, Burovski, Peterson, Weckesser, Bright, {van der Walt}, Brett,
  Wilson, Millman, Mayorov, Nelson, Jones, Kern, Larson, Carey, Polat, Feng,
  Moore, {VanderPlas}, Laxalde, Perktold, Cimrman, Henriksen, Quintero, Harris,
  Archibald, Ribeiro, Pedregosa, {van Mulbregt}, \& {SciPy 1.0
  Contributors}}]{2020SciPy-NMeth}
Virtanen, P., Gommers, R., Oliphant, T.~E., {et~al.} 2020, Nature Methods, 17,
  261, \dodoi{10.1038/s41592-019-0686-2}

\bibitem[{{Wurster} {et~al.}(2016){Wurster}, {Price}, \& {Bate}}]{Wurster2016}
{Wurster}, J., {Price}, D.~J., \& {Bate}, M.~R. 2016, \mnras, 457, 1037,
  \dodoi{10.1093/mnras/stw013}

\bibitem[{{Yang} {et~al.}(2021){Yang}, {Sakai}, {Zhang}, {Murillo}, {Zhang},
  {Higuchi}, {Zeng}, {L{\'o}pez-Sepulcre}, {Yamamoto}, {Lefloch}, {Bouvier},
  {Ceccarelli}, {Hirota}, {Imai}, {Oya}, {Sakai}, \& {Watanabe}}]{Yang2021}
{Yang}, Y.-L., {Sakai}, N., {Zhang}, Y., {et~al.} 2021, \apj, 910, 20,
  \dodoi{10.3847/1538-4357/abdfd6}

\bibitem[{{Yorke} \& {Bodenheimer}(1999)}]{YB1999}
{Yorke}, H.~W., \& {Bodenheimer}, P. 1999, \apj, 525, 330,
  \dodoi{10.1086/307867}

\bibitem[{{Yorke} {et~al.}(1993){Yorke}, {Bodenheimer}, \&
  {Laughlin}}]{YBL1993}
{Yorke}, H.~W., {Bodenheimer}, P., \& {Laughlin}, G. 1993, \apj, 411, 274,
  \dodoi{10.1086/172827}

\bibitem[{{Yorke} {et~al.}(1995){Yorke}, {Bodenheimer}, \&
  {Laughlin}}]{YBL1995}
---. 1995, \apj, 443, 199, \dodoi{10.1086/175514}

\bibitem[{{Yu} {et~al.}(2018){Yu}, {Ho}, {Barth}, \& {Li}}]{Yu2018}
{Yu}, S.-Y., {Ho}, L.~C., {Barth}, A.~J., \& {Li}, Z.-Y. 2018, \apj, 862, 13,
  \dodoi{10.3847/1538-4357/aacb25}

\bibitem[{{Yu} {et~al.}(2019){Yu}, {Ho}, \& {Zhu}}]{Yu2019}
{Yu}, S.-Y., {Ho}, L.~C., \& {Zhu}, Z. 2019, \apj, 877, 100,
  \dodoi{10.3847/1538-4357/ab1d65}

\bibitem[{{Zhao} {et~al.}(2016){Zhao}, {Caselli}, {Li}, {Krasnopolsky},
  {Shang}, \& {Nakamura}}]{Zhao2016}
{Zhao}, B., {Caselli}, P., {Li}, Z.-Y., {et~al.} 2016, \mnras, 460, 2050,
  \dodoi{10.1093/mnras/stw1124}

\bibitem[{{Zhao} {et~al.}(2011){Zhao}, {Li}, {Nakamura}, {Krasnopolsky}, \&
  {Shang}}]{ZLNKS2011}
{Zhao}, B., {Li}, Z.-Y., {Nakamura}, F., {Krasnopolsky}, R., \& {Shang}, H.
  2011, \apj, 742, 10, \dodoi{10.1088/0004-637X/742/1/10}

\end{thebibliography}
\bibliographystyle{aasjournal}
  
\appendix

\section{Notes on cell integrations}
Many of the data presented here are in the form of integrals of quantities inside a volume of a computational cell.

Due to grid staggering we performed our integrals in cell-centered manner with the help of interpolation.

The center of the cell is numbered as $(i,j,k)$, with the centers of the cell faces at points called $(i-\h,j,k)$, $(i+\h,j,k)$ and so on,
with $i$, $j$, and $k$ in the $r$, $\theta$, and $\phi$ directions.
This amounts, in terms of the variables of the Zeus codes, to labeling the $b$-grid with integer numbers and the $a$-grid with half-integer numbers,
We keep to this convention while acknowledging that the converse notation (integers for the $a$-grid) is also used for other works.

Using this convention, when we compute integrals in a cell-centered manner, we are integrating inside a computational volume $V_{i,j,k}$, centered at the point $(i,j,k)$, and extending in the $r,\theta,\phi$ directions to the range $[i-\h, i+\h],[j-\h,j+\h],[k-\h,k+\h]$,
A face-centered point is located at a place such as $(i-\h,j,k)$, and the corresponding volume $V_{i-\h,j,k}$ extends to the range $[i-1, i],[j-\h,j+\h],[k-\h,k+\h]$.

We can start from the torque term
\[N_\mathrm{prs}=-\int \frac{\partial P}{\partial\phi}\,dV\ \]
where the pressure $P$ can be seen as either of the thermal pressure $p$, or the total pressure $p+p_\mathrm{mag}$.
We assume that the pressure $P$ is given to us defined at cell centers $(i,j,k)$ and that we want to compute the cell-centered integral over the volume
$V_{i,j,k}$.

\subsection{Point value approximation}
While our $r$ and $\theta$ grids are non-uniform, our $\phi$ grid is uniform.
That allows a quick computation of the 
point-value of the partial derivative at the cell center as
\[
\left.\frac{\partial P}{\partial\phi}\right|_{i,j,k}
=
\frac{P(i,j,k+1)-P(i,j,k-1)}{2\,\Delta\phi} + O(\Delta\phi^\mathrm{order})\ ,
\]
using the familiar centered-difference formula for the partial derivative.
The Point-value approximation consists in
just multiplying this point value at the center of $V$ by the volume of $V$.
This neglects the variations of the force inside of $V$. Such approximation is expected to give results in the correct order of magnitude nearly everywhere, with some inaccuracy in regions where the force undergoes rapid variations.

\section{Disk plane in model H}\label{sec:Append_diskplane}

\restartappendixnumbering

\begin{figure}[htb]
	\centering
	\includegraphics[width=0.8\columnwidth]{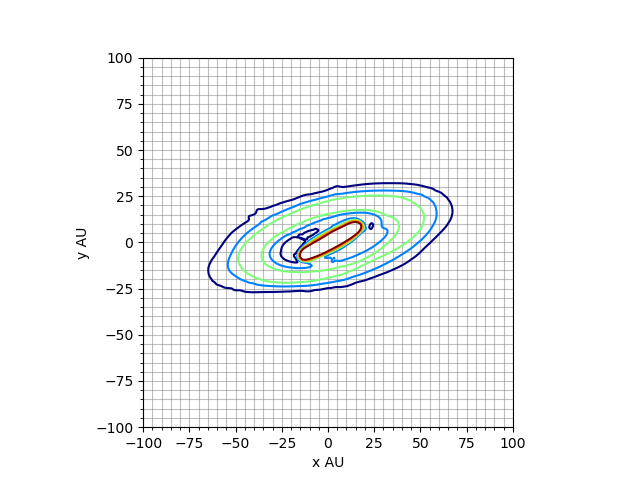}
	\caption{Contours of column density in model H at $154.3\kyr$ viewed at a position angle of ($88{\degree}$, $288{\degree}$).}\label{fig:Hcolden_cntr}
\end{figure}

Edge-on view of column density provides the tilt angle of disk. In model H the disk has noticeable tilting. Figure \ref{fig:Hcolden_cntr} shows the contours of column density in model H at the same frame with Figure \ref{fig:Hcolden}. The viewing angle ($\theta,\phi$) is an input parameter for PERSPECTIVE to provide column density. When the contour of largest column density (e.g. red curve) becomes the thinnest, its corresponding viewing angle is chosen to find the normal direction to disk plane. The slope of red contour is denoted by ${\rm arctan}(y,x)$. Thus the normal direction to disk plane is given by $(\theta+{\rm arctan}(y,x),\phi)$.

\section{2DDFT}\label{sec:Appendx_2ddft}
To characterize spiral structure, we implement 2DDFT to extract the mode and pitch angle of spirals. The Fourier component ${\rm AMP}(p,m)$ can be derived by the Fourier transform of column density $N(r,\varphi)$ as

\begin{equation}
    {\rm AMP}(p,m)=\frac{1}{2\pi}\int^{2\pi}_0\int^\infty_{-\infty}N({\rm ln}r,\varphi)\exp{[-i(p{\rm ln}r+m\varphi)]}\intd ({\rm ln}r)\intd\varphi.
\end{equation}

For discretely sampled $N$, the normalized discrete Fourier transform for ${\rm AMP}(p,m)$ can be obtained as

\begin{align}
    {\rm AMP}(p,m) & = \frac{1}{\Sigma^M_{n=1} N_n}\int_{{\rm ln}r_{\rm in}}^{{\rm ln}r_{\rm out}}\int_\pi^{-\pi}\nonumber\\
    & \sum^M_{n=1} N_n(r_n,\varphi_n)\delta({\rm ln}r-{\rm ln}r_n)\delta(\varphi-\varphi_n)\exp{[-i(m\varphi+p{\rm ln}r)]}\intd\varphi\intd({\rm ln}r)\\
    & = \frac{1}{\Sigma^M_{n=1} N_n}\sum^M_{n=1} N_n(r_n,\varphi_n)\exp{[-i(m\varphi_n+p{\rm ln}r_n)]},
\end{align}
where $N_n(r_n,\varphi_n)$ is the column density of the $n-$th grid cell at $(r_n,\varphi_n)$, $r_{\rm in}$ and $r_{\rm out}$ are the inner and outer boundary of spiral structure, and $M$ is the number of grids between $r_{\rm in}$ and $r_{\rm out}$. The mode in $\varphi$-direction is sampled as $m=[1,2,3,4]$ and in $r$-direction $p$ is $100$-equally sampled in the range of $(-20,20)$. The pitch angle $\alpha$ can be written in a function of dominant mode $(p,m)$ as $\alpha={\rm arctan}(|\frac{m}{p}|)$. 

\end{document}